% Something to deal with sub-sub-sections

\def\unlockat{\catcode`\@=11}

\unlockat

% Sets global section and proposition numbers to zero
\global\newcount\secno \global\secno=0
\global\newcount\prono \global\prono=0
%%%%%%%%% Big sections
\def\newsec#1{\global\advance\secno by1\message{(\the\secno. #1)}
\global\subsecno=0\global\subsubsecno=0
\global\deno=0\global\teno=0
\eqnres@t\noindent
{\titlefont\the\secno. #1}
\writetoca{{\bf\secsym} { #1}}\par\nobreak\medskip\nobreak}
\global\newcount\subsecno \global\subsecno=0
%%%%%%%%%%%%%%%%Subsections %%%%%%%%%%%%%%%%%%%%%%%%%%%%%%%%%%%%%%%
\def\subsec#1{\global\advance\subsecno
by1\message{(\secsym\the\subsecno. #1)}
\ifnum\lastpenalty>9000\else\bigbreak\fi
\global\subsubsecno=0
\global\deno=0
\global\teno=0
%\eqnres@t
\noindent{\bf\secsym\the\subsecno. #1}
\writetoca{\bf \string\quad {\secsym\the\subsecno.} {\it  #1}}
\par\nobreak\medskip\nobreak}
\global\newcount\subsubsecno \global\subsubsecno=0
%%%%%%%%%%%%%%%%%%%%Subsubsections %%%%%%%%%%%%%%%%%%%%%%%%%%%%%
\def\subsubsec#1{\global\advance\subsubsecno by1
\message{(\secsym\the\subsecno.\the\subsubsecno. #1)}
\ifnum\lastpenalty>9000\else\bigbreak\fi
\noindent\quad{\bf \secsym\the\subsecno.\the\subsubsecno.}{\ \sl \ #1}
\writetoca{\string\qquad\bf { \secsym\the\subsecno.\the\subsubsecno.}{\sl  \ #1}}
\par\nobreak\medskip\nobreak}
%%% Definition

\global\newcount\deno \global\deno=0
\def\de#1{\global\advance\deno by1
\message{(\bf Definition\quad\secsym\the\subsecno.\the\deno #1)}
\ifnum\lastpenalty>9000\else\bigbreak\fi
\noindent{\bf Definition\quad\secsym\the\subsecno.\the\deno}{#1}
\writetoca{\string\qquad{\secsym\the\subsecno.\the\deno}{#1}}}
%%% Proposition

\global\newcount\prono \global\prono=0
\def\pro#1{\global\advance\prono by1
\message{(\bf Proposition\quad\secsym\the\subsecno.\the\prono %#1
)}
\ifnum\lastpenalty>9000\else\bigbreak\fi
\noindent{\bf Proposition\quad%\secsym\the\subsecno.
\the\prono\quad}{\ninepoint #1}
%\writetoca{\string\qquad{\secsym\the\subsecno.\the\prono}{#1}
}
%%% Theorem

\global\newcount\teno \global\prono=0
\def\te#1{\global\advance\teno by1
\message{(\bf Theorem\quad\secsym\the\subsecno.\the\teno #1)}
\ifnum\lastpenalty>9000\else\bigbreak\fi
\noindent{\bf Theorem\quad\secsym\the\subsecno.\the\teno}{#1}
\writetoca{\string\qquad{\secsym\the\subsecno.\the\teno}{#1}}}
%%%%%%%%%%%%
\def\subsubseclab#1{\DefWarn#1\xdef #1{\noexpand\hyperref{}{subsubsection}%
{\secsym\the\subsecno.\the\subsubsecno}%
{\secsym\the\subsecno.\the\subsubsecno}}%
\writedef{#1\leftbracket#1}\wrlabeL{#1=#1}}

\def\unredoffs{} \def\redoffs{\voffset=-.40truein\hoffset=-.40truein}
\def\speclscape{}

\newbox\leftpage \newdimen\fullhsize \newdimen\hstitle \newdimen\hsbody
\tolerance=1000\hfuzz=2pt

\catcode`\@=11
\def\bigans{b }
\def\answ{b }

\ifx\answ\bigans\message{(This will come out unreduced.}
\magnification=1200\unredoffs\baselineskip=16pt plus 2pt minus 1pt
\hsbody=\hsize \hstitle=\hsize

\else\message{(This will be reduced.} \let\l@r=L
\magnification=1200\baselineskip=16pt plus 2pt minus 1pt \vsize=7truein
\redoffs \hstitle=8truein\hsbody=4.75truein\fullhsize=10truein\hsize=\hsbody
\output={\ifnum\pageno=0

   \shipout\vbox{{\hsize\fullhsize\makeheadline}
     \hbox to \fullhsize{\hfill\pagebody\hfill}}\advancepageno
   \else
   \almostshipout{\leftline{\vbox{\pagebody\makefootline}}}\advancepageno
   \fi}
\def\almostshipout#1{\if L\l@r \count1=1 \message{[\the\count0.\the\count1]}
       \global\setbox\leftpage=#1 \global\let\l@r=R
  \else \count1=2
   \shipout\vbox{\speclscape{\hsize\fullhsize\makeheadline}
       \hbox to\fullhsize{\box\leftpage\hfil#1}}  \global\let\l@r=L\fi}
\fi

\newcount\yearltd\yearltd=\year\advance\yearltd by -2000

\def\Title#1#2{%\nopagenumbers
\abstractfont\hsize=\hstitle\rightline{#1}%
\vskip 5pt\centerline{\titlefont #2}\abstractfont\vskip .5in\pageno=0}
%

%&%

\def\draftmode{\message{ DRAFTMODE }\def\draftdate{{\rm preliminary draft:
\number\month/\number\day/\number\yearltd\ \ \hourmin}}%

\writelabels\baselineskip=20pt plus 2pt minus 2pt
  {\count255=\time\divide\count255 by 60 \xdef\hourmin{\number\count255}
   \multiply\count255 by-60\advance\count255 by\time
   \xdef\hourmin{\hourmin:\ifnum\count255<10 0\fi\the\count255}}}

\def\nolabels{\def\wrlabeL##1{}\def\eqlabeL##1{}\def\reflabeL##1{}}
\def\writelabels{\def\wrlabeL##1{\leavevmode\vadjust{\rlap{\smash%
{\line{{\escapechar=` \hfill\rlap{\sevenrm\hskip.03in\string##1}}}}}}}%
\def\eqlabeL##1{{\escapechar-1\rlap{\sevenrm\hskip.05in\string##1}}}%
\def\reflabeL##1{\noexpand\llap{\noexpand\sevenrm\string\string\string##1}}}
\nolabels

\global\newcount\secno \global\secno=0
\global\newcount\meqno
\global\meqno=1
\def\eqnres@t{\xdef\secsym{\the\secno.}\global\meqno=1
\bigbreak\bigskip}
\def\sequentialequations{\def\eqnres@t{\bigbreak}}
\def\appendix#1#2{\vfill\eject\global\meqno=1\global\subsecno=0\xdef\secsym{\hbox{#1.}}
\bigbreak\bigskip\noindent{\bf Appendix #1. #2}\message{(#1. #2)}
\writetoca{Appendix {#1.} {#2}}\par\nobreak\medskip\nobreak}

\def\eqnn#1{\xdef #1{(\secsym\the\meqno)}\writedef{#1\leftbracket#1}%
\global\advance\meqno by1\wrlabeL#1}
\def\eqna#1{\xdef #1##1{\hbox{$(\secsym\the\meqno##1)$}}
\writedef{#1\numbersign1\leftbracket#1{\numbersign1}}%
\global\advance\meqno by1\wrlabeL{#1$\{\}$}}
\def\eqn#1#2{\xdef #1{(\secsym\the\meqno)}\writedef{#1\leftbracket#1}%
\global\advance\meqno by1$$#2\eqno#1\eqlabeL#1$$}

\newskip\footskip\footskip14pt plus 1pt minus 1pt

\def\footnotefont{\ninepoint}\def\f@t#1{\footnotefont #1\@foot}
\def\f@@t{\baselineskip\footskip\bgroup\footnotefont\aftergroup\@foot\let\next}
\setbox\strutbox=\hbox{\vrule height9.5pt depth4.5pt width0pt}
\global\newcount\ftno \global\ftno=0
\def\foot{\global\advance\ftno by1\footnote{$^{\the\ftno}$}}

\newwrite\ftfile
\def\footend{\def\foot{\global\advance\ftno by1\chardef\wfile=\ftfile
$^{\the\ftno}$\ifnum\ftno=1\immediate\openout\ftfile=foots.tmp\fi%
\immediate\write\ftfile{\noexpand\smallskip%
\noexpand\item{f\the\ftno:\ }\pctsign}\findarg}%
\def\footatend{\vfill\eject\immediate\closeout\ftfile{\parindent=20pt
\centerline{\bf Footnotes}\nobreak\bigskip\input foots.tmp }}}
\def\footatend{}

\global\newcount\refno \global\refno=1
\newwrite\rfile
\def\ref{[\the\refno]\nref}
\def\nref#1{\xdef#1{[\the\refno]}\writedef{#1\leftbracket#1}%
\ifnum\refno=1\immediate\openout\rfile=refs.tmp\fi \global\advance\refno
by1\chardef\wfile=\rfile\immediate \write\rfile{\noexpand\item{#1\
}\reflabeL{#1\hskip.31in}\pctsign}\findarg}

\def\findarg#1#{\begingroup\obeylines\newlinechar=`\^^M\pass@rg}
{\obeylines\gdef\pass@rg#1{\writ@line\relax #1^^M\hbox{}^^M}%
\gdef\writ@line#1^^M{\expandafter\toks0\expandafter{\striprel@x #1}%
\edef\next{\the\toks0}\ifx\next\em@rk\let\next=\endgroup\else\ifx\next\empty%
\else\immediate\write\wfile{\the\toks0}\fi\let\next=\writ@line\fi\next\relax}}
\def\striprel@x#1{} \def\em@rk{\hbox{}}
\def\lref{\begingroup\obeylines\lr@f}
\def\lr@f#1#2{\gdef#1{\ref#1{#2}}\endgroup\unskip}
\def\semi{;\hfil\break}
\def\addref#1{\immediate\write\rfile{\noexpand\item{}#1}}

\def\startrefs#1{\immediate\openout\rfile=refs.tmp\refno=#1}
\def\xref{\expandafter\xr@f}\def\xr@f[#1]{#1}
\def\refs#1{\count255=1[\r@fs #1{\hbox{}}]}
\def\r@fs#1{\ifx\und@fined#1\message{reflabel \string#1 is undefined.}%
\nref#1{need to supply reference \string#1.}\fi%
\vphantom{\hphantom{#1}}\edef\next{#1}\ifx\next\em@rk\def\next{}%
\else\ifx\next#1\ifodd\count255\relax\xref#1\count255=0\fi%
\else#1\count255=1\fi\let\next=\r@fs\fi\next}

%%%%%%%TABLE OF CONTENT%%%%%%%%%%%%%

\def\writetoc{\immediate\openout\tfile=nikmac.tmp
    \def\writetoca##1{{\edef\next{\write\tfile{\noindent  ##1
    \string\leaderfill {\noexpand\number\pageno} \par}}\next}}}

%       and this lists table of contents on second pass

%
\catcode`\@=12 % at signs are no longer letters
%
%   Unpleasantness in calling in abstract and title fonts
\edef\tfontsize{\ifx\answ\bigans scaled\magstep3\else scaled\magstep4\fi}
\font\titlerm=cmr10 \tfontsize \font\titlerms=cmr7 \tfontsize
\font\titlermss=cmr5 \tfontsize \font\titlei=cmmi10 \tfontsize
\font\titleis=cmmi7 \tfontsize \font\titleiss=cmmi5 \tfontsize
\font\titlesy=cmsy10 \tfontsize \font\titlesys=cmsy7 \tfontsize
\font\titlesyss=cmsy5 \tfontsize \font\titleit=cmti10 \tfontsize
\skewchar\titlei='177 \skewchar\titleis='177 \skewchar\titleiss='177
\skewchar\titlesy='60 \skewchar\titlesys='60 \skewchar\titlesyss='60
\def\titlefont{\def\rm{\fam0\titlerm}% switch to title font
\textfont0=\titlerm \scriptfont0=\titlerms \scriptscriptfont0=\titlermss
\textfont1=\titlei \scriptfont1=\titleis \scriptscriptfont1=\titleiss
\textfont2=\titlesy \scriptfont2=\titlesys \scriptscriptfont2=\titlesyss
\textfont\itfam=\titleit
\def\it{\fam\itfam\titleit}\rm}
\font\authorfont=cmcsc10 \ifx\answ\bigans\else scaled\magstep1\fi
\ifx\answ\bigans\def\abstractfont{\tenpoint}\else \font\abssl=cmsl10 scaled
\magstep1 \font\absrm=cmr10 scaled\magstep1 \font\absrms=cmr7
scaled\magstep1 \font\absrmss=cmr5 scaled\magstep1 \font\absi=cmmi10
scaled\magstep1 \font\absis=cmmi7 scaled\magstep1 \font\absiss=cmmi5
scaled\magstep1 \font\abssy=cmsy10 scaled\magstep1 \font\abssys=cmsy7
scaled\magstep1 \font\abssyss=cmsy5 scaled\magstep1 \font\absbf=cmbx10
scaled\magstep1 \skewchar\absi='177 \skewchar\absis='177
\skewchar\absiss='177 \skewchar\abssy='60 \skewchar\abssys='60
\skewchar\abssyss='60
\def\abstractfont{\def\rm{\fam0\absrm}% switch to abstract font
\textfont0=\absrm \scriptfont0=\absrms \scriptscriptfont0=\absrmss
\textfont1=\absi \scriptfont1=\absis \scriptscriptfont1=\absiss
\textfont2=\abssy \scriptfont2=\abssys \scriptscriptfont2=\abssyss
\textfont\itfam=\bigit \def\it{\fam\itfam\bigit}\def\footnotefont{\tenpoint}%
\textfont\slfam=\abssl \def\sl{\fam\slfam\abssl}%
\textfont\bffam=\absbf \def\bf{\fam\bffam\absbf}\rm}\fi
\def\tenpoint{\def\rm{\fam0\tenrm}% switch back to 10-point type
\textfont0=\tenrm \scriptfont0=\sevenrm \scriptscriptfont0=\fiverm
\textfont1=\teni  \scriptfont1=\seveni  \scriptscriptfont1=\fivei
\textfont2=\tensy \scriptfont2=\sevensy \scriptscriptfont2=\fivesy
\textfont\itfam=\tenit \def\it{\fam\itfam\tenit}\def\footnotefont{\ninepoint}%
\textfont\bffam=\tenbf
\def\bf{\fam\bffam\tenbf}\def\sl{\fam\slfam\tensl}\rm}
\font\ninerm=cmr9 \font\sixrm=cmr6 \font\ninei=cmmi9 \font\sixi=cmmi6
\font\ninesy=cmsy9 \font\sixsy=cmsy6 \font\ninebf=cmbx9 \font\nineit=cmti9
\font\ninesl=cmsl9 \skewchar\ninei='177 \skewchar\sixi='177
\skewchar\ninesy='60 \skewchar\sixsy='60
\def\ninepoint{\def\rm{\fam0\ninerm}% switch to footnote font
\textfont0=\ninerm \scriptfont0=\sixrm \scriptscriptfont0=\fiverm
\textfont1=\ninei \scriptfont1=\sixi \scriptscriptfont1=\fivei
\textfont2=\ninesy \scriptfont2=\sixsy \scriptscriptfont2=\fivesy
\textfont\itfam=\ninei \def\it{\fam\itfam\nineit}\def\sl{\fam\slfam\ninesl}%
\textfont\bffam=\ninebf \def\bf{\fam\bffam\ninebf}\rm}
%
%---------------------------------------------------------------------
%

\hyphenation{anom-aly anom-alies coun-ter-term coun-ter-terms}
\def\inv{^{\raise.15ex\hbox{${\scriptscriptstyle -}$}\kern-.05em 1}}

\def\Dsl{\,\raise.15ex\hbox{/}\mkern-13.5mu D} 
%this one can be subscripted
\def\dsl{\raise.15ex\hbox{/}\kern-.57em\partial}

\def\tr#1{\, {\rm tr}\, \left( #1 \right)}

 %pound sterling
\def\lspace{\ifx\answ\bigans{}\else\qquad\fi}
\def\lbspace{\ifx\answ\bigans{}\else\hskip-.2in\fi} % $$\lbspace...$$
\def\boxeqn#1{\vcenter{\vbox{\hrule\hbox{\vrule\kern3pt\vbox{\kern3pt
     \hbox{${\displaystyle #1}$}\kern3pt}\kern3pt\vrule}
    }}}
\def\mbox#1#2{\vcenter{\hrule \hbox{\vrule height#2in
         \kern#1in \vrule} \hrule}}  %e.g. \mbox{.1}{.1}

%%%%%%%%%%%%%%%%%%%%%%%%

\newwrite\ffile\global\newcount\figno \global\figno=1
\def\nfig#1{\xdef#1{fig.~\the\figno}%
\writedef{#1\leftbracket fig.\noexpand~\the\figno}%
\ifnum\figno=1\immediate\openout\ffile=figs.tmp\fi\chardef\wfile=\ffile%
\immediate\write\ffile{\noexpand\medskip\noexpand\item{Fig.\ \the\figno. }
\reflabeL{#1\hskip.55in}\pctsign}\global\advance\figno by1\findarg}
\def\xfig{\expandafter\xf@g}
\def\xf@g fig.\penalty\@M\ {}
\def\figs#1{figs.~\f@gs #1{\hbox{}}}
\def\f@gs#1{\edef\next{#1}\ifx\next\em@rk\def\next{}\else
\ifx\next#1\xfig #1\else#1\fi\let\next=\f@gs\fi\next}
\newwrite\lfile
{\escapechar-1\xdef\pctsign{\string\%}\xdef\leftbracket{\string\{}
\xdef\rightbracket{\string\}}\xdef\numbersign{\string\#}}

\def\writestop{\def\writestoppt{\immediate\write\lfile{\string\pageno%
\the\pageno\string\startrefs\leftbracket\the\refno\rightbracket%
\string\def\string\secsym\leftbracket\secsym\rightbracket%
\string\secno\the\secno\string\meqno\the\meqno}\immediate\closeout\lfile}}
\def\writestoppt{}\def\writedef#1{}
\def\seclab#1{\xdef #1{\the\secno}\writedef{#1\leftbracket#1}\wrlabeL{#1=#1}}
\def\subseclab#1{\xdef #1{\secsym\the\subsecno}%
\writedef{#1\leftbracket#1}\wrlabeL{#1=#1}}
\newwrite\tfile \def\writetoca#1{}
\def\leaderfill{\leaders\hbox to 1em{\hss.\hss}\hfill}

%%%%%% Tildes and hats%%%%%%%%%%%%%%%

\def\tilde{\widetilde}
\def\bar{\overline}
\def\hat{\widehat}

%%%%%%%%%%%%% Cech %%%%%%%%%%%%
\def\cech{${\rm C}^{\kern-6pt \vbox{\hbox{$\scriptscriptstyle\vee$}\kern2.5pt}}${\rm ech}}
\def\Cech{${\sl C}^{\kern-6pt \vbox{\hbox{$\scriptscriptstyle\vee$}\kern2.5pt}}${\sl ech}}

%%%%%%%%%%%% Greek %%%%%%%%%%%%

\def\d{{\delta}}

\def\e{{\epsilon}}

\def\l{{\lambda}}

\def\n{{\nu}}

\def\s{{\sigma}}

\def\vt{{\vartheta}}

%%%%% Non-terms %%%%%

%%%%%%%%%%%% Derivatives  %%%%%%%%%%%
\def\p{\partial}

\def\inv{^{\raise.15ex\hbox{${\scriptscriptstyle -}$}\kern-.05em 1}}

\def\Dsl{\,\raise.15ex\hbox{/}\mkern-13.5mu D}
%this one can be subscripted
\def\dsl{\raise.15ex\hbox{/}\kern-.57em\partial}

%%%%%%%%%%% bold %%%%%%%%%%%%%%

\def\bA{{\bf A}}

\def\bB{{\bf B}}

\def\bC{{\bf C}}

\def\bL{{\bf L}}

\def\bm{{\bf m}}
\def\bN{{\bf N}}

\def\bQ{{\bf Q}}

\def\bS{{\bf S}}
\def\bT{{\bf T}}
\def\bt{{\bf t}}

\def\bZ{{\bf Z}}

%%%%%%%%%%% letters with bar %%%%%%%%

%%%%%%%%% underlined letters %%%%%%%%%%%

%%%%%%%%%%%%%%%  Rublenye bukvy   %%%%%%%%%%%%%%%%%
\def\IB{\relax\hbox{$\inbar\kern-.3em{\rm B}$}}

\def\ID{\relax\hbox{$\inbar\kern-.3em{\rm D}$}}
\def\IE{\relax\hbox{$\inbar\kern-.3em{\rm E}$}}
\def\IF{\relax\hbox{$\inbar\kern-.3em{\rm F}$}}
\def\IG{\relax\hbox{$\inbar\kern-.3em{\rm G}$}}
\def\IGa{\relax\hbox{${\rm I}\kern-.18em\Gamma$}}
\def\IH{\relax{\rm I\kern-.18em H}}
\def\IK{\relax{\rm I\kern-.18em K}}
\def\IL{\relax{\rm I\kern-.18em L}}
\def\IP{\relax{\rm I\kern-.18em P}}
\def\II{\relax{\rm I\kern-.18em I}}

\def\ndt{{\noindent}}

%%%%%%%% Calligraphic letters  %%%%%%%%%%%%%

\def\CH{{\cal H}}

\def\CN{{\cal N}}
\def\CO{{\cal O}}

\def\CQ{{\cal Q}}

\def\CS{{\cal S}}

%%%%%%%%%%% letters with bar %%%%%%%%

%%%%%%%%%% Math symbols %%%%%%%%%%%%%

\def\HH{{\bf H}}

\def\lime{{\rm Lim}_{\kern -16pt \vbox{\kern6pt\hbox{$\scriptstyle{\e \to 0}$}}}}

\def\naiveq{\qquad =^{\kern-12pt \vbox{\hbox{$\scriptscriptstyle{\rm naive}$}\kern5pt}} \qquad}

%
%---------------------------------------------------------------------
%

\hyphenation{anom-aly anom-alies coun-ter-term coun-ter-terms}
\def\tr{\, {\rm tr}\, }

 %pound sterling
\def\lspace{\ifx\answ\bigans{}\else\qquad\fi}
\def\lbspace{\ifx\answ\bigans{}\else\hskip-.2in\fi} % $$\lbspace...$$
\def\boxeqn#1{\vcenter{\vbox{\hrule\hbox{\vrule\kern3pt\vbox{\kern3pt
      \hbox{${\displaystyle #1}$}\kern3pt}\kern3pt\vrule}\hrule}}}
\def\mbox#1#2{\vcenter{\hrule \hbox{\vrule height#2in
          \kern#1in \vrule} \hrule}}  %e.g. \mbox{.1}{.1}
%   matters of taste
%\def\tilde{\widetilde} \def\bar{\overline} \def\hat{\widehat}
%
% some sample definitions

\def\darr#1{\raise1.5ex\hbox{$\leftrightarrow$}\mkern-16.5mu #1}
 %pound sterling

\def\half{{\textstyle{1\over2}}} %puts a small half in a displayed eqn
\def\roughly#1{\raise.3ex\hbox{$#1$\kern-.75em\lower1ex\hbox{$\sim$}}}

%%%%%%%%%%%%%% Lie algebras %%%%%%%%%%%

\def\inbar{\,\vrule height1.5ex width.4pt depth0pt}

%%%%%%%%%%% Macros for boxes %%%%%%%%%%%
\def\boxit#1{\vbox{\hrule\hbox{\vrule\kern8pt
\vbox{\hbox{\kern8pt}\hbox{\vbox{#1}}\hbox{\kern8pt}}
\kern8pt\vrule}\hrule}}
\def\mathboxit#1{\vbox{\hrule\hbox{\vrule\kern8pt\vbox{\kern8pt
\hbox{$\displaystyle #1$}\kern8pt}\kern8pt\vrule}\hrule}}

%%%%%%%%%%%%%%%%%%%%%%%%

%%%%% Latex %%%%%%%
\def\mathcal#1{{\cal #1}}

\def\frac#1#2{{{ #1 }\over{ #2 }}}
\def\frac1#1{{1\over{#1}}}

%%%%%%%%%%%%%%%%%%
%%%% Journals %%%%%%%%%
\def\np#1#2#3{Nucl.~Phys.~{\bB#1}(#2) #3}
\def\cmp#1#2#3{Comm.~Math.~Phys.~{\bf #1}(#2) #3}
\def\plb#1#2#3{Phys.~Lett.~{\bB#1} (#2) #3}
\def\jhep#1#2#3{JHEP~{\bB#1} (#3) #2}

%REFERENCES

\lref\bass{N.~Nekrasov, S.~Shatashvili, ``Supersymmetric vacua and quantum integrability,'' to appear}
\lref\nazarov{M.~Nazarov, V.~Tarasov,``On irreducibility of tensor products of Yangian modules associated with skew Young diagrams'', arXiv:math/0012039, Duke~Math.~J. {\bf 112} (2002), 343-378\semi
M.~Nazarov, V.~Tarasov, ``On Irreducibility of Tensor Products of Yangian Modules'',
arXiv:q-alg/9712004,
Internat.~Math.~Research~Notices (1998) 125-150\semi
M.~Nazarov, V.~Tarasov, ``Representations of Yangians with Gelfand-Zetlin Bases'',
arXiv:q-alg/9502008,  J.~Reine~Angew.~Math. {\bf 496} (1998) 181-212\semi
M.~Nazarov, V.~Tarasov, ``Yangians and Gelfand-Zetlin bases'', arXiv:hep-th/9302102, Publ.~Res.~Inst.~Math.~Sci. Kyoto {\bf 30} (1994) 459-478}
\lref\korepin{V.~Korepin, N.~Bogolyubov, A.~Izergin, ``Quantum Inverse Scattering Method and Correlation Functions'', Cambridge  Monographs on Mathematical Physics, Cambridge University Press, 1997}
\lref\dubrovintt{B.~Dubrovin, ``Geometry and Integrability of Topological-Antitopological Fusion'', arXiv:hep-th/9206037,   \cmp{152}{1993}{539-564} }

\lref\cutr{For the current situation see, ``Integrability in String and Gauge Theory'', Utrecht, August, 2008}
\lref\mz{J.~Minahan, K.~Zarembo, ``The Bethe-Ansatz for ${\CN}=4$ Super Yang-Mills, ''
arXiv:hep-th/0212208 , \jhep{0303}{2003}{013}}
\lref\vafacr{C.~Vafa, ``Topological Mirrors and Quantum Rings,'' in, {\it Essays on Mirror Manifolds}, ed. S.-T.~Yau (Intl.Press, 1992)}
\lref\vc{S.~Cecotti, C.~Vafa, ``Topological Anti-Topological Fusion'', \np{367}{1991}{359-461}}
\lref\gnell{A.~Gorsky, N.~Nekrasov, ``Elliptic Calogero-Moser system from two dimensional current algebra'', arXiv: hep-th/9401021}
\lref\gnthd{A.~Gorsky, N.~Nekrasov, ``Relativistic Calogero-Moser model as gauged WZW theory'', \np{436}{1995}{582-608}, arXiv:hep-th/9401017}
\lref\mnph{J.A.~Minahan, A.P.~Polychronakos, ``Interacting Fermion Systems from Two Dimensional QCD,'' 
\plb{326}{1994}{288-294}, arXiv:hep-th/9309044
    \semi
~~~~``Equivalence of Two Dimensional QCD and the $c=1$ Matrix Model'', \plb{312}{1993}{155-165}, arXiv:hep-th/9303153   \semi
~~~~ ``Integrable Systems for Particles with Internal Degrees of Freedom'',
    \plb{302}{1993}{265-270},
    arXiv:hep-th/9206046}
\lref\gntd{A.~Gorsky, N.~Nekrasov, ``Hamiltonian systems of Calogero type and two dimensional Yang-Mills theory'', arXiv:hep-th/9304047, \np{414}{1994}{213-238}}
\lref\gstwo{A.~Gerasimov, S.L.~Shatashvili, ``Two-dimensional gauge theories and quantum integrable systems'', arXiv:0711.1472, in, {\it "From Hodge Theory to Integrability and TQFT: tt*-geometry"}, pp. 239-262,
R.~Donagi and K.~Wendland, Eds., Proc. of Symposia in
Pure Mathematics Vol. 78, American Mathematical Society
Providence, Rhode Island.  }
\lref\yangyang{C.~N.~Yang, C.~P.~Yang, J.~Math.~Phys~{\bf 10} (1969) 1115}
\lref\ll{E.~H.~Lieb, W.~Liniger, Phys.~Rev.~{\bf 130} (1963) 1605}
\lref\toda{M.~Toda, Prog.~Theor.~Phys.~Suppl.~{\bf 45} (1970) 174}
\lref\todaba{M.~Opper, ``Analytical Solution of the Classical Bethe-Ansatz Equation for the Toda Chain'', Phys. Lett. {\bA}, 112 (1985) 201-203}
\lref\nazarov{M.~Nazarov, V.~Tarasov,``On irreducibility of tensor products of Yangian modules associated with skew Young diagrams'', arXiv:math/0012039, Duke~Math.~J. {\bf 112} (2002), 343-378\semi
M.~Nazarov, V.~Tarasov, ``On Irreducibility of Tensor Products of Yangian Modules'',
arXiv:q-alg/9712004,
Internat.~Math.~Research~Notices (1998) 125-150\semi
M.~Nazarov, V.~Tarasov, ``Representations of Yangians with Gelfand-Zetlin Bases'',
arXiv:q-alg/9502008,  J.~Reine~Angew.~Math. {\bf 496} (1998) 181-212\semi
M.~Nazarov, V.~Tarasov, ``Yangians and Gelfand-Zetlin bases'', arXiv:hep-th/9302102, Publ.~Res.~Inst.~Math.~Sci. Kyoto {\bf 30} (1994) 459-478}
\lref\baxter{R.~Baxter, ``Exactly solved models in statistical mechanics'', London, Academic Press, 1982}
\lref\dubrovintt{B.~Dubrovin, ``Geometry and Integrability of Topological-Antitopological Fusion'', arXiv:hep-th/9206037,   \cmp{152}{1993}{539-564} }
\lref\nekhol{N.~Nekrasov, ``Holomorphic bundles and many-body systems'', arXiv:hep-th/9503157, \cmp{180}{1996}{587-604} }
\lref\GE{S.~Katz, A.~Klemm, C.~Vafa,
``Geometric Engineering of Quantum Field Theories'',  arXiv:hep-th/9609239}
\lref\NF{N.~Nekrasov,~
``Five Dimensional Gauge
Theories and Relativistic Integrable Systems'',  arXiv:hep-th/9609219}
\lref\NL{A.~Lawrence, N.~Nekrasov,~
``Instanton sums and five-dimensional gauge theories'',~  arXiv:hep-th/9706025}
\lref\MaxManin{M.~Kontsevich, Yu.~Manin, ``Gromov-Witten classes, quantum cohomology, and enumerative geometry '', arXiv:hep-th/9402147}
\lref\WittenTwoD{E.~Witten, ``Two dimensional gauge theory revisited'', arXiv:hep-th/9204083}
\lref\gmmm{A.~Gorsky, A.~Marshakov, A.~Mironov, A.~Morozov, ``${\CN}=2$ Supersymmetric QCD and Integrable Spin Chains: Rational Case $N_f < 2N_c$'', arXiv: hep-th/9603140, \plb{380}{1996}{75-80}\semi
A.~Gorsky, S.~Gukov, A.~Mironov,``SUSY field theories, integrable systems and their stringy/brane origin -- II'', arXiv:hep-th/9710239,
\np{518}{1998}{689-713}\semi
A.~Gorsky, S.~Gukov, A.~Mironov,
``Multiscale ${\CN}=2$ SUSY field theories, integrable systems and their stringy/brane origin -- I
'', arXiv:hep-th/9707120, \np{517}{1998}{409-461} \semi
R.~Boels, J.~de Boer,
``Classical Spin Chains and Exact Three-dimensional Superpotentials'',
arXiv:hep-th/0411110}

\lref\gkmmm{A.~Gorsky, I.~Krichever, A.~Marshakov, A.~Mironov, A.~Morozov}

\lref\niksw{N.~Nekrasov, ``Seiberg-Witten prepotential
from instanton calculus,'' arXiv:hep-th/0206161,  arXiv:~hep-th/0306211}

\lref\nikokounkov{N.~Nekrasov, A.~Okounkov, ``Seiberg-Witten theory and random partitions,''
 arXiv:~hep-th/0306238}

\lref\kolyaf{N.~Reshetikhin, ``The functional equation method in the theory of exactly soluble quantum systems'', ZhETF {\bf 84} (1983), 1190-1201 (in Russian) Sov. Phys. JETP {\bf 57} (1983), 691-696 (English Transl.)}

\lref\baxteriii{R.J.~Baxter, ``Partition Function of the Eight-Vertex Lattice Model'' , Ann. Phys. {\bf 70} (1972) 193-228\semi
``One Dimensional Anisotropic Heisenberg Chain'', Ann. Phys. {\bf 70} (1972) 323-337\semi ``Eight-Vertex Model in Lattice Statistics and One Dimensional Anisotropic Heisenberg Chain I, II, III'', Ann. Phys. {\bf 76} (1973) 1-24, 25-47, 48-71}

\lref\baeref{N.~Yu.~Reshetikhin,  ``Integrable models of quantum
one-dimensional magnets with
$O(n)$- and
$Sp(2k)$-symmetries'', (Russian) Teoret. Mat. Fiz. 63 (1985), no. 3,
347--366}
\lref\baerefi{
E.~Ogievetsky, N.~Yu.~Reshetikhin, P.~Wiegmann, ``The principal
chiral field in two dimensions on classical Lie algebras. The
Bethe-ansatz solution and factorized theory of scattering'',
\np{280}{1987}{no. 1, 45-96}\semi
N.~Yu.~Reshetikhin, P.~Wiegmann, ``Towards the classification
of completely integrable quantum field theories (the Bethe-ansatz
associated with Dynkin diagrams and their automorphisms)'',
\plb{189}{1987}{no. 1-2, 125-131}}
\lref\baerefiii{
E.~Ogievetsky, P.~Wiegmann,
``Factorized $S$-matrix and the Bethe
ansatz for simple Lie groups'', \plb{168}{1986}{no. 4,
360--366}}
\lref\baekr{
A.~N.~Kirillov,
N.~Yu.~Reshetikhin, ``Representations of Yangians
and multiplicities of the inclusion of the irreducible components
of the tensor product of representations of simple Lie algebras'' ,
(Russian) Zap. Nauchn. Sem. Leningrad. Otdel. Mat. Inst. Steklov.
(LOMI) 160 (1987), Anal. Teor. Chisel i Teor. Funktsii. 8,
211--221, 301; translation in J.~Soviet~Math. {\bf 52} (1990), no. 3,
3156--3164}

\lref\gerasimovshatashvili{A.~Gerasimov, S.L.~Shatashvili, ``Higgs Bundles, Gauge Theories and Quantum Groups'', \cmp{277}{2008}{323-367}, arXiv:hep-th/0609024}

\lref\nikthesis{N.~Nekrasov, ``Four dimensional holomorphic theories'',
PhD. thesis, Princeton University, 1996, UMI-9701221}

\lref\HH{A.~Hanany, K.~Hori, ``Branes and ${\CN}=2$ Theories in Two Dimensions'', arXiv:hep-th/9707192}

%%%%% Old refs  %%%%%%

\lref\blzh{A. Belavin, V. Zakharov, ``Yang-Mills Equations as inverse
scattering
problem''Phys. Lett. B73, (1978) 53}
\lref\bost{L. Alvarez-Gaume, J.B. Bost , G. Moore, P. Nelson, C.
Vafa,
``Bosonization on higher genus Riemann surfaces,''
Commun.Math.Phys.112:503,1987}
\lref\agmv{L.~Alvarez-Gaum\'e,
C.~Gomez, G.~Moore,
and C.~Vafa, ``Strings in the Operator Formalism,''
\np{303}{1988}{455}}
\lref\atiyah{M.~Atiyah, ``Green's Functions for
Self-Dual Four-Manifolds,'' Adv. Math. Suppl.
{\bf 7A} (1981)129}

\lref\donagi{R.~Y.~ Donagi,
``Seiberg-Witten integrable systems'',
alg-geom/9705010 }

\lref\AHS{M.~F.~Atiyah, N.~ Hitchin and I.~M.~Singer, ``Self-Duality in
Four-Dimensional
Riemannian Geometry", Proc. Royal Soc. (London) {\bf A362} (1978)
425-461.}
\lref\fmlies{M.~F.~Atiyah and I.~M.~Singer,
``The index of elliptic operators IV,'' Ann. Math. {\bf 93}(1968)119}
\lref\BlThlgt{M.~ Blau and G.~ Thompson, ``Lectures on 2d Gauge
Theories: Topological Aspects and Path
Integral Techniques", Presented at the
Summer School in High Energy Physics and
Cosmology, Trieste, Italy, 14 Jun - 30 Jul
1993, hep-th/9310144.}
\lref\bpz{A.A.~Belavin, A.M.~Polyakov, A.B.~Zamolodchikov,
``Infinite conformal symmetry in two-dimensional quantum
field theory,'' \npb{241}{1984}{333}}
\lref\braam{P.J.~Braam, A.~Maciocia, and A.~Todorov,
``Instanton moduli as a novel map from tori to
K3-surfaces,'' Inv.~Math.~{\bf 108} (1992) 419}
\lref\CMR{ For a review, see
S.~Cordes, G.~Moore, and S.~Ramgoolam,
`` Lectures on 2D Yang Mills theory, Equivariant
Cohomology, and Topological String Theory,''
Lectures presented at the 1994 Les Houches Summer School
 ``Fluctuating Geometries in Statistical Mechanics and Field
Theory.''
and at the Trieste 1994 Spring school on superstrings.
hep-th/9411210, or see http://xxx.lanl.gov/lh94}
\lref\dnld{S.~K.~Donaldson, ``Anti self-dual Yang-Mills
connections over complex  algebraic surfaces and stable
vector bundles,'' Proc. Lond. Math. Soc,
{\bf 50} (1985)1}

\lref\DoKro{S.K.~ Donaldson and P.B.~ Kronheimer,
{\it The Geometry of Four-Manifolds},
Clarendon Press, Oxford, 1990.}
\lref\donii{
S. Donaldson, Duke Math. J. , {\bf 54} (1987) 231. }

\lref\gerasimov{A.~Gerasimov, ``Localization in
GWZW and Verlinde formula,'' hepth/9305090}

\lref\gottsh{L.~G\"ottsche, Math.~Ann. {\bf 286} (1990)193}
\lref\GrHa{P.~ Griffiths and J.~ Harris, {\it Principles of
Algebraic
geometry},
p. 445, J.Wiley and Sons, 1978. }

\lref\hipg{N. Hitchin, ``Polygons and gravitons,''
Math. Proc. Camb. Phil. Soc, (1979){\bf 85} 465}
\lref\hitchin{N.~Hitchin, ``Stable bundles and integrable systems'', Duke Math
{\bf 54}  (1987),91-114}
\lref\hid{N.~Hitchin, ``The self-duality equations on a Riemann surface'',
Proc. London Math. Soc. {\bf 55} (1987) 59-126 }
\lref\hklr{N.~Hitchin, A.~Karlhede, U.~Lindstrom, and M.~Rocek,
``Hyperkahler metrics and supersymmetry,''
\cmp{108}{1987}{ 535}}
\lref\hirz{F. Hirzebruch and T. Hofer, Math. Ann. 286 (1990)255}
\lref\btverlinde{M.~ Blau, G.~ Thomson,
``Derivation of the Verlinde Formula from Chern-Simons Theory and the
$G/G$
   model'', \np{408}{1993}{345-390} }
\lref\kronheimer{P.~Kronheimer, ``The construction of ALE spaces as
hyper-kahler quotients,'' J. Diff. Geom. {\bf 28}1989)665}
\lref\kricm{P. Kronheimer, ``Embedded surfaces in
4-manifolds,'' Proc. Int. Cong. of
Math. (Kyoto 1990) ed. I. Satake, Tokyo, 1991}

\lref\krmw{P.~Kronheimer and T.~Mrowka,
``Gauge theories for embedded surfaces I,''
Topology {\bf 32} (1993) 773,
``Gauge theories for embedded surfaces II,''
preprint.}
%%%%%%%%%%%%%%%%%%%%%%%%%%%%%%%%%%%%%%%%%%%%%%%%%%%%%%
%%%%%%%%%%%%% Kirwan %%%%%%%%%%%%%%%%%%%%%%%%%%%%%%%%
\lref\kirwan{F.~Kirwan, ``Cohomology of quotients in symplectic
and algebraic geometry'', Math. Notes, Princeton University Press, 1985}
%%%%%%%%%%%%%%%%%% Avatar %%%%%%%%%%%%%%%%%%%%%%%%%%%
\lref\avatar{A. Losev, G. Moore, N. Nekrasov, S. Shatashvili,
``Four-Dimensional Avatars of 2D RCFT,''
hep-th/9509151, Nucl.Phys.Proc.Suppl.46:130-145,1996 }
\lref\cocycle{A. Losev, G. Moore, N. Nekrasov, S. Shatashvili,
``Central Extensions of Gauge Groups Revisited,''
hep-th/9511185.}
%%%%%%%%%%%%%%%%%%%%%%%%%%%%%%%%%%%%%%%%%%
\lref\maciocia{A. Maciocia, ``Metrics on the moduli
spaces of instantons over Euclidean 4-Space,''
Commun. Math. Phys. {\bf 135}(1991) , 467}
%%%%%%%%%%%%%%%%%%%%%%%%%%%%%%%%%%%%%%%%%%%%%%
\lref\mickold{J. Mickelsson, CMP, 97 (1985) 361.}
\lref\mick{J. Mickelsson, ``Kac-Moody groups,
topology of the Dirac determinant bundle and
fermionization,'' Commun. Math. Phys., {\bf 110} (1987) 173.}
%%%%%%%%%%%%%%%%%%%%%%%%%%%%%%%%%%%%%%%%%%%%%%%%%
\lref\milnor{J. Milnor, ``A unique decomposition
theorem for 3-manifolds,'' Amer. Jour. Math, (1961) 1}
%%%%%%%%%%%%%%%%%%%%%%%%%%%%5
\lref\taming{G. Moore and N. Seiberg,
``Taming the conformal zoo,'' Phys. Lett.
{\bf 220 B} (1989) 422}
%%%%%%%%%%%%%%%%%%%%%%%%%%%%%%%%%%%%%%%%%%%%%%%%%%%
\lref\nair{V.P.Nair, ``K\"ahler-Chern-Simons Theory'', hep-th/9110042}
\lref\ns{V.P. Nair and Jeremy Schiff,
``Kahler Chern Simons theory and symmetries of
antiselfdual equations'' Nucl.Phys.B371:329-352,1992;
``A Kahler Chern-Simons theory and quantization of the
moduli of antiselfdual instantons,''
Phys.Lett.B246:423-429,1990,
``Topological gauge theory and twistors,''
Phys.Lett.B233:343,1989}
\lref\ogvf{H. Ooguri and C. Vafa, ``Self-Duality
and ${\CN}=2$ String Magic,'' Mod.Phys.Lett. {\bf A5} (1990) 1389-1398;
``Geometry
of${\CN}=2$ Strings,'' \np{361}{1991}{469-518}}
%%%%%%%%%%%%%%%% Park %%%%%%%%%%%%%%%%%%%%%%%%%%%%
\lref\park{J.-S. Park, ``Holomorphic Yang-Mills theory on compact
Kahler
manifolds,'' hep-th/9305095; \np{423}{1994}{559};
J.-S.~ Park, ``$N=2$ Topological Yang-Mills Theory on Compact
K\"ahler
Surfaces", Commun. Math, Phys. {\bf 163} (1994) 113;
S. Hyun and J.-S.~ Park, ``$N=2$ Topological Yang-Mills Theories and Donaldson
Polynomials", hep-th/9404009}
\lref\parki{S. Hyun and J.-S. Park,
``Holomorphic Yang-Mills Theory and Variation
of the Donaldson Invariants,'' hep-th/9503036}
\lref\dpark{J.-S.~Park, ``Monads and D-instantons'', hep-th/9612096}
%%%%%%%%%%%%%%%%%%%%%%%%%%%%%%%%%%%%%%%%%%%%%%%%%%%%%%%%%%
\lref\pohl{Pohlmeyer, \cmp{72}{1980}{37}}
%%%%%%%%%%%%%%%%%%%%%%%%%%%%%%%%%%%%%%%
\lref\pwf{A.M. Polyakov and P.B. Wiegmann,
\plb{131}{1983}{121}}
\lref\clash{
A.~Losev, G.~Moore, N.~Nekrasov, S.~Shatashvili,
`` Chiral  Lagrangians, Anomalies, Supersymmetry, and Holomorphy'', Nucl.Phys.
{\bf B} 484(1997) 196-222, hep-th/9606082 }
\lref\givental{A.B.~Givental,
``Equivariant Gromov - Witten Invariants'',
alg-geom/9603021}
%%%%%%%%%%%%%%%%%%%%%%%%%%%%%%%%%%%%
\lref\prseg{Pressley and Segal, Loop Groups}
\lref\rade{J. Rade, ``Singular Yang-Mills fields. Local
theory I. '' J. reine ang. Math. , {\bf 452}(1994)111; {\it ibid}
{\bf 456}(1994)197; ``Singular Yang-Mills
fields-global theory,'' Intl. J. of Math. {\bf 5}(1994)491.}
\lref\segal{G. Segal, The definition of CFT}
\lref\sen{A. Sen,
hep-th/9402032, Dyon-Monopole bound states, selfdual harmonic
forms on the multimonopole moduli space and $SL(2,Z)$
invariance,'' }
\lref\shatashi{S. Shatashvili,
Theor. and Math. Physics, 71, 1987, p. 366}
\lref\thooft{G. 't Hooft , ``A property of electric and
magnetic flux in nonabelian gauge theories,''
Nucl.Phys.B153:141,1979}
\lref\vafa{C. Vafa, ``Conformal theories and punctured
surfaces,'' Phys.Lett.199B:195,1987 }
%%%%%%%%%%%%%%%% Verlinde %%%%%%%%%%%%%%%%%%%%
\lref\vrlsq{E. Verlinde and H. Verlinde,
``Conformal Field Theory and Geometric Quantization,''
in {\it Strings'89},Proceedings
of the Trieste Spring School on Superstrings,
3-14 April 1989, M. Green, et. al. Eds. World
Scientific, 1990}

\lref\mwxllvrld{E. Verlinde, ``Global Aspects of
Electric-Magnetic Duality,'' hep-th/9506011}

%%%%%%%%%%%%%%%%%%%%%%%%%%%%%%%%%%%%%%%%%%
\lref\wrdhd{R. Ward, Nucl. Phys. {\bf B236}(1984)381}
\lref\ward{Ward and Wells, {\it Twistor Geometry and
Field Theory}, CUP }

%%%%%%%%%%%%%%%%% Witten %%%%%%%%%%%%%%%%%%%%%%%%%%%%%%%%

\lref\WitDonagi{R.~ Donagi, E.~ Witten,
``Supersymmetric Yang-Mills Theory and
Integrable Systems'', hep-th/9510101, Nucl.Phys.{\bf B}460 (1996) 299-334}
\lref\Witfeb{E.~ Witten, ``Supersymmetric Yang-Mills Theory On A
Four-Manifold,'' J. Math. Phys. {\bf 35} (1994) 5101.}

\lref\Witgrav{E.~ Witten, ``Topological Gravity'', Phys.Lett.206B:601, 1988}
\lref\witaffl{I. ~ Affleck, J.A.~ Harvey and E.~ Witten,
    ``Instantons and (Super)Symmetry Breaking
    in $2+1$ Dimensions'', Nucl. Phys. {\bf B}206 (1982) 413}
\lref\wittabl{E.~ Witten,  ``On $S$-Duality in Abelian Gauge Theory,''
hep-th/9505186; Selecta Mathematica {\bf 1} (1995) 383}
\lref\wittgr{E.~ Witten, ``The Verlinde Algebra And The Cohomology Of
The Grassmannian'',  hep-th/9312104}
\lref\wittenwzw{E. Witten, ``Nonabelian bosonization in
two dimensions,'' Commun. Math. Phys. {\bf 92} (1984)455 }
\lref\witgrsm{E. Witten, ``Quantum field theory,
grassmannians and algebraic curves,'' Commun.Math.Phys.113:529,1988}
\lref\wittjones{E. Witten, ``Quantum field theory and the Jones
polynomial,'' Commun.  Math. Phys., 121 (1989) 351. }
\lref\witttft{E.~ Witten, ``Topological Quantum Field Theory",
Commun. Math. Phys. {\bf 117} (1988) 353.}
\lref\wittmon{E.~ Witten, ``Monopoles and Four-Manifolds'', hep-th/9411102}
\lref\Witdgt{ E.~ Witten, ``On Quantum gauge theories in two
dimensions,''
Commun. Math. Phys. {\bf  141}  (1991) 153\semi
 ``Two dimensional gauge
theories revisited'', J. Geom. Phys. 9 (1992) 303-368}
\lref\Witgenus{E.~ Witten, ``Elliptic Genera and Quantum Field Theory'',
Comm. Math. Phys. 109(1987) 525. }
\lref\issues{A.~Losev, N.~Nekrasov, S.~Shatashvili, ``Issues in topological gauge theory'', \np{534}{1998}{549-611}, arXiv:hep-th/9711108}
\lref\OldZT{E. Witten, ``New Issues in Manifolds of SU(3) Holonomy,''
{\it Nucl. Phys.} {\bf B268} (1986) 79 \semi
J. Distler and B. Greene, ``Aspects of (2,0) String Compactifications,''
{\it Nucl. Phys.} {\bf B304} (1988) 1 \semi
B. Greene, ``Superconformal Compactifications in Weighted Projective
Space,'' \cmp{130}{1990}{335}}
\lref\nikfive{N.~Nekrasov, ``Five dimensional gauge theories and relativistic integrable systems'',
\np{531}{1998}{323-344},
arXiv:hep-th/9609219}
\lref\bagger{E.~ Witten and J. Bagger, \plb{115}{1982}{202}}
\lref\witcurrent{E.~ Witten,``Global Aspects of Current Algebra'',
Nucl.Phys.B223 (1983) 422\semi
``Current Algebra, Baryons and Quark Confinement'', Nucl.Phys. B223 (1993)
433}
\lref\Wittreiman{S.B. Treiman,
E. Witten, R. Jackiw, B. Zumino, ``Current Algebra and
Anomalies'', Singapore, Singapore: World Scientific ( 1985) }
\lref\Witgravanom{L. Alvarez-Gaume, E.~ Witten, ``Gravitational Anomalies'',
Nucl.Phys.B234:269,1984. }

\lref\CHSW{P.~Candelas, G.~Horowitz, A.~Strominger and E.~Witten,
``Vacuum Configurations for Superstrings,'' {\it Nucl. Phys.} {\bf
B258} (1985) 46.}

\lref\AandB{E.~Witten, in ``Proceedings of the Conference on Mirror Symmetry",
MSRI (1991).}

\lref\phases{E.~Witten, ``Phases of ${\CN}=2$ Theories in Two Dimensions",
Nucl. Phys. {\bf B403} (1993) 159, hep-th/9301042}
\lref\WitKachru{S.~Kachru and E.~Witten, ``Computing The Complete Massless
Spectrum Of A Landau-Ginzburg Orbifold,''
Nucl. Phys. {\bf B407} (1993) 637, hep-th/9307038}

\lref\WitMin{E.~Witten,
``On the Landau-Ginzburg Description of N=2 Minimal Models,''
IASSNS-HEP-93/10, hep-th/9304026.}

\lref\wittentft{E.~ Witten, ``Topological Quantum Field Theory",
Commun. Math. Phys. {\bf 117} (1988) 353.}
\lref\Witdgt{ E.~ Witten, ``On Quantum gauge theories in two
dimensions,''
Commun. Math. Phys. {\bf  141}  (1991) 153.}
\lref\Witfeb{E.~ Witten, ``Supersymmetric Yang-Mills Theory On A
Four-Manifold,'' J. Math. Phys. {\bf 35} (1994) 5101.}
\lref\WitIntroCoh{E.~ Witten, ``Introduction to Cohomological Field
Theories",
Lectures at Workshop on Topological Methods in Physics, Trieste,
Italy,
Jun 11-25, 1990, Int. J. Mod. Phys. {\bf A6} (1991) 2775.}
\lref\wittabl{E. Witten,  ``On S-Duality in Abelian Gauge Theory,''
hep-th/9505186}

\lref\seiken{K. Intriligator, N. Seiberg,
``Mirror Symmetry in Three Dimensional Gauge Theories'',
hep-th/9607207, Phys.Lett. B387 (1996) 513}
\lref\douglas{M.R. Douglas, ``Enhanced Gauge
Symmetry in M(atrix) Theory,'' hep-th/9612126}
\lref\hs{J.~A.~Harvey and A.~Strominger,
``The heterotic string is a soliton,''
hep-th/9504047}
\lref\HarveyMooreBPS{ J.~A.~Harvey, G.~Moore,
``On the algebras of BPS states'', hep-th/9609017}
\lref\zt{O.~Aharony, M.~Berkooz, N.~Seiberg, ``Light-cone description
of $(2,0)$ supersonformal theories in six dimensions'', hep-th/9712117}
\lref\senstringduality{A.~Sen, ``String-String Duality Conjecture In Six Dimensions And
Charged Solitonic Strings'',  hep-th/9504027}
%%%%%%%%%%%%%%%%%%%%%%%%%%%%%%%%%%%%%%%%%%%%%%%%%%%%%%
%%%%%%%%%%%%% Nakajima %%%%%%%%%%%%%%%%%%%%%%%%%%%%%%%
\lref\KN{P.~Kronheimer and H.~Nakajima,  ``Yang-Mills instantons
on ALE gravitational instantons,''  Math. Ann.
{\bf 288}(1990)263}
\lref\nakajimahom{H.~Nakajima, ``Homology of moduli
spaces of instantons on ALE Spaces. I'' J. Diff. Geom.
{\bf 40}(1990) 105; ``Instantons on ALE spaces,
quiver varieties, and Kac-Moody algebras,'' Duke. Math. J. {\bf 76} (1994)
365-416\semi
``Gauge theory on resolutions of simple singularities
and affine Lie algebras,'' Inter. Math. Res. Notices (1994), 61-74}
\lref\nakheis{H.~Nakajima, ``Heisenberg algebra and Hilbert schemes of
points on
projective surfaces ,'' alg-geom/9507012\semi
``Lectures on Hilbert schemes of points on surfaces'', H.~Nakajima's
homepage}
\lref\vw{C.~Vafa, E.~Witten, ``A strong coupling test of $S$-duality'',
\npb{431}{1994}{3-77}}
\lref\grojn{I.~Grojnowski, ``Instantons and
affine algebras I: the Hilbert scheme and
vertex operators,''
alg-geom/9506020}

\lref\gyangian{H.~Nakajima, ``Quiver varieties and finite dimensional representations of quantum affine algebras,'' 
arXiv:math/9912158}
\lref\mvyangian{M.~Varagnolo,  ``Quiver varieties and Yangians,'' arXiv:math/0005277}

\lref\gibrych{G.~Gibbons, P.~Rychenkova, ``Hyperkahler quotient construction
of BPS Monopole moduli space'', hep-th/9608085}
\lref\dvafa{
C.~Vafa, ``Instantons on D-branes'', hep-th/9512078, \npb{463}{1996}{435-442}}
\lref\atbott{M.~Atiyah, R.~Bott, ``The Moment Map And
Equivariant Cohomology'', Topology {\bf 23} (1984) 1-28}
\lref\atbotti{M.~Atiyah, R.~Bott, 
``The Yang-Mills Equations Over
Riemann Surfaces'', Philosophical Transactions of the Royal Society of London. Series {\bA}, Mathematical and Physical Sciences, Volume 308, Issue 1505, pp. 523-615}

\lref\higgs{G.~Moore, N.~Nekrasov, S.~Shatashvili,
``Integration over the Higgs branches'', \cmp{209}{2000}{97-121}, arXiv:hep-th/9712241}

%%%%%%%%%%%%%%%%%%%%%%%%%%%
\Title{ \vbox{\baselineskip12pt
\hbox{IHES-P/08/59}
\hbox{TCD-MATH-09-05}
\hbox{HMI-09-02}
\hbox{NSF-KITP-09-12}
}}
{\vbox{
\bigskip
\bigskip
 \centerline{QUANTUM INTEGRABILITY}
 \bigskip
 \centerline{AND}
 \bigskip
 \centerline{SUPERSYMMETRIC VACUA }
}}
\medskip
\centerline{\authorfont Nikita A.
Nekrasov\footnote{$^{a}$}{On leave of absence
from ITEP, Moscow, Russia}$^{,1}$,
and Samson L. Shatashvili$^{1,2,3}$}
\vskip 0.5cm
\centerline{\it $^{1}$ Institut des Hautes Etudes Scientifiques,
Bures-sur-Yvette, France}
\centerline{\it $^{2}$ Hamilton Mathematical Institute, Trinity College,
Dublin 2, Ireland}
\centerline{\it $^{3}$ School of Mathematics, Trinity College, Dublin 2, Ireland}
\vskip 0.1cm

\bigskip
\ndt
{\ninepoint
Supersymmetric vacua of two dimensional ${\CN}=4$ gauge theories with matter, softly broken
by the twisted masses down to
${\CN}=2$, are shown to be in 
one-to-one correspondence with the eigenstates of integrable spin chain Hamiltonians.  Examples include:
the Heisenberg
$SU(2)$ $XXX$ spin chain which is mapped to the two dimensional $U(N)$ theory with fundamental
hypermultiplets, the $XXZ$ spin chain which is mapped  to the analogous three dimensional super-Yang-Mills theory compactified on a circle,  the $XYZ$ spin chain and eight-vertex model which are related to the four dimensional theory compactified on ${\bT}^{2}$. 
A consequence of our correspondence is the isomorphism of the quantum cohomology ring of various quiver varieties, such as $T^{*}{\rm Gr}(N,L)$ and the ring of quantum integrals of motion of various spin chains. 
The correspondence extends to any spin group, representations, boundary conditions, and inhomogeneity, it includes Sinh-Gordon and non-linear Schr\"odinger models as well as the dynamical spin chains like Hubbard model. We give the gauge-theoretic interpretation of Drinfeld polynomials and Baxter operators. The two-sphere compactifications of the four dimensional ${\CN}=2$ theories  lead to the instanton corrected Bethe equations. We suggest the Yangian, quantum affine, and elliptic algebras are a completely novel kind of symmetry of the (collections of the) interacting quantum field theories.}

\centerline{\it To Prof. T.~Eguchi on the occasion of his 60th anniversary}

\newsec{Gauge theories and integrable systems}

The dynamics of gauge theory is a subject of long history and the ever growing importance.

In the last fifteen years or so it has become clear that the gauge theory dynamics in the vacuum sector is related to that of quantum many-body systems. A classic example is the equivalence of the
pure Yang-Mills theory with gauge group $U(N)$ in two dimensons to the system of $N$ free non-relativistic fermions on a circle. The same theory embeds as a supersymmetric vacuum sector of a (deformation of) ${\CN}=2$ super-Yang-Mills theory in two dimensions.

A bit less trivial example found in \higgs\ is that the vacuum sector of a
certain supersymmetric two dimensional $U(N)$ gauge theory with massive adjoint matter is described by the solutions of Bethe ansatz equations for the quantum Nonlinear Schr\"odinger equation (NLS) in the $N$-particle sector. The model of \higgs\ describes the $U(1)$-equivariant intersection theory on the moduli space of solutions to Hitchin's equations \hitchin, just as the pure Yang-Mills
theory describes
 the intersection theory on moduli space of flat connections on a two dimensional Riemann surface. This subject was revived in \gerasimovshatashvili,\gstwo,  by showing that the natural interpretation of the results of \higgs\ is in terms of the equivalence of the vacua of the $U(N)$ Yang-Mills-Higgs theory in a sense of \gerasimovshatashvili\ and the energy eigenstates of the $N$-particle Yang system, i.e. a system of $N$ non-relativistic particles on a circle with delta-function interaction.  Furthermore, \gerasimovshatashvili,\gstwo\ suggested that such a correspondence should be a general property of a larger class of supersymmetric gauge theories in various spacetime dimensions.

Prior to \higgs\  a different connection to spin systems with long-range interaction appeared in two dimensional pure Yang-Mills theory with massive matter \gntd, 
\mnph. Three dimensional lift of latter gauge theory describes relativistic interacting particles \gnthd,
while four dimensional theories lead to elliptic generalizations
\gnell.

In this paper we formulate precisely the correspondence between the two dimensional ${\CN}=2$ supersymmetric gauge theories and quantum integrable systems in a very general setup. The ${\CN}=2$ supersymmetric theories have rich algebraic structure surviving quantum corrections \vc. In particular, there is a distinguished class of operators $\left( {\CO}_{A} \right)$, which commute with some of the nilpotent supercharges $\CQ$ of the supersymmetry algebra. They have no singularities in their operator product expansion and, when considered up to the $\CQ$-commutators,  form a (super)commutative ring, called the chiral ring \vc,\vafacr. The supersymmetric vacua of the theory form a representation of that ring. The space of supersymmetric vacua is thus naturally identified with the space of states of a quantum integrable system, whose Hamiltonians are the generators of the chiral ring.
The duality states that the spectrum of the quantum Hamiltonians coincides with the spectrum of the chiral ring.
The nontrivial result of this paper is that arguably all quantum integrable lattice models from the integrable systems textbooks correspond in this fashion to the ${\CN}=2$
supersymmetric {\it gauge} theories, essentially also from the
(different) textbooks. More precisely, the gauge theories which correspond to the integrable spin chains and their limits (the non-linear Schr\"odinger equation and other systems encountered in \higgs,\gerasimovshatashvili,\gstwo\ being particular large spin limits thereof) are the softly broken
${\CN}=4$ theories. It is quite important that we are dealing here with the gauge theories, rather then the general $(2,2)$ models, since it is in the gauge theory context that the equations describing the supersymmetric vacua can be identified with Bethe equations of the integrable world.

At this point we should clarify a possible confusion about the r\^ole
 of integrable systems in the description of the dynamics of supersymmetric gauge theories.

It is known that the low energy
dynamics of the four dimensional ${\CN}=2$ supersymmetric
gauge theories is governed by the classical algebraic integrable systems \WitDonagi. Moreover, the natural
gauge theories lead to integrable systems of Hitchin type, which are equivalent to many-body systems \nekhol\ and conjecturally to spin chains
\gmmm.

We emphasize, however, that the correspondence between
the  gauge theories and integrable models we discuss in the present paper and in \higgs,\gerasimovshatashvili,\gstwo\ is of a different nature.  The low energy effective theory in four dimensions is described by the classical algebraic integrable systems of type \WitDonagi, 
while the vacuum states we discuss presently are mapped to the quantum eigenstates of a different, quantum integrable system\foot{Another possible source of confusion is the emergence of the
Bethe ansatz and the spin chains in the ${\CN}=4$ supersymmetric
gauge theory in four dimensions.
In the work \mz\ and its further developments \cutr\  the anomalous dimensions of  local operators of the ${\CN}=4$ supersymmetric Yang-Mills theory
are shown (to a certain loop order in perturbation theory) to be the eigenvalues of
some spin chain Hamiltonian. The
gauge theory is studied in the 't Hooft large $N$ limit. In our story the gauge theory has less supersymmetry, $N$ is finite, and the operators we consider are from the chiral ring, i.e. their conformal dimensions are not corrected
quantum mechanically. Our goal is
to determine their vacuum expectation values.}.

The gauge theories we shall study in two dimensions, as well as their string theory realizations,  have a natural lift to three and four dimensions,
while keeping the same number of supersymmetries, modulo certain
anomalies. Indeed, the ${\CN}=2$
super-Yang-Mills theory in two dimensions is a dimensional reduction of the ${\CN}=1$ four dimensional Yang-Mills theory
(this fact is useful in
the superspace formulation
of the theory \phases).
Instead of the dimensional reduction one can take the compactification on a two dimensional torus. That way the theory will look macroscopically two dimensional, but the effective dynamics will be different due to the
contributions of the Kaluza-Klein modes (the early examples of these
corrections in the analogous
compactifications from five to four dimensions can be found in \nikfive). This is seen, for example, in the geometry of the (classical) moduli space of vacua, which
is compact for the theory obtained by compactification from four to two dimensions (it is isomorphic to the moduli space $Bun_{G}$ of holomorphic $G_{\bC}$-bundles  on elliptic curve), and is non-compact in the dimensionally reduced theory. Quantum mechanically, though, the geometry of the moduli space of vacua is more complicated, in particular it will acquire many components. The twisted superpotential is a meromorphic function on the moduli space.
We show that the critical points of this function determine the Bethe roots of the anisotropic spin chain, the $XYZ$ magnet. Its $XXZ$ limit will
be mapped to the
three dimensional gauge theory compactified on a circle. We thus get a satisfying picture of the elliptic, trigonometric, and rational theories  corresponding to the four dimensional, three dimensional  and the two dimensional theories respectively.

The duality between the gauge theories and the quantum integrable systems we established in this paper can be used to enrich both subjects. For example, the notions of {\it special coordinates, topological/anti-topological fusion} \vc, and so on have not been appreciated so far in the world of quantum integrable systems.

In this note we shall mostly discuss the example which relates the $XXX$ spin chain for the $SU(2)$ group, and the ${\CN}=4$ two dimensional theory with the gauge group $U(N)$ and $L$ fundamental hypermultiplet, whose supersymmetry is broken down to ${\CN}=2$ by the choice of the twisted masses. In some limit the
theory reduces to the supersymmetric sigma model on the noncompact hyperk\"ahler manifold, that of the cotangent bundle to the Grassmanian Gr$(N,L)$ of the $N$-dimensional complex planes in ${\bC}^{L}$. 
Our main statement then maps the equivariant quantum cohomology algebra of $T^{*}{\rm Gr}(N,L)$
to the algebra of quantum integrals
of motion of the $XXX_{1/2}$ spin chain.

$\underline{\rm A\ longer\ version.}$ This note is a shortened version of \bass.
In 
\bass\ we give the precise microscopic description of the matter sector, superpotential and twisted superpotential of the theories under consideration. We also explain how one lifts these theories to three and four dimensions. We then compute the twisted effective superpotential ${\tilde W}^{\rm eff}({\s})$ on the Coulomb branch  for the all our models. We then derive the exact equations describing quantum-mechanical supersymmetric ground states:
\eqn\main{
{\exp} \left( {\p {\tilde W}^{\rm eff}}(\sigma) \over {\p \s}^{i}  \right) =  1}
We present several examples, and remind the connection to quantum cohomology of various homogeneous spaces, like Grassmanians and flag varieties.
We then discuss the theories with
${\CN}=4$ supersymmetric matter content softly broken down to ${\CN}=2$ by the twisted masses. In this case the equations
\main\
are identified with the Bethe equations of the dual quantum integrable systems.
Also \bass\ reviews the methods used to solve exactly spin chains and related quantum integrable systems, like the eight vertex model, Hubbard model, Gaudin model, non-linear Schr\"odinger system and so on.
Finally, after all these preparations we formulate the duality dictionary between the gauge theories and spin chains. We show that the Bethe eigenvectors in quantum integrable systems correspond to the supersymmetric ground states in gauge theory. We identify the so-called Yang-Yang (YY) function \yangyang\ of quantum integrable system with the effective twisted superpotential
${\tilde W}$ of the gauge theory, thus showing the universal character of the observations in \higgs,\gerasimovshatashvili,\gstwo. The vacuum equation \main\  then coincides with the Bethe equation, while the Hamiltonians correspond to the chiral ring observables. 
We discuss the naturalness of the gauge theories which we map to quantum integrable systems. We show that the pattern of the twisted masses which at first appears highly fine tuned is in fact the generic pattern of twisted masses
compatible with the superpotential of the microscopic theory.
We show that the gauge theories with $U(N)$ group map to periodic spin chains, the gauge theories with
$SO(N), Sp(N)$ gauge groups are mapped to open spin chains with particular boundary conditions. We show that the $A, D, E$-type, as well as the supergroup spin chains, with various representations at
the spin sites correspond to quiver gauge theories with the
$\times_{i} U(N_{i})$ gauge groups. 
Furthermore, \bass\ provides the string theory construction of some of these theories. The string theory point of view makes some of the tools of the algebraic Bethe ansatz 
\ref\stf{L.D.~Faddeev, E.~Sklyanin, L.~Takhtajan,  ``Quantum inverse problem method'', Theor. Math. Phys. {\bf 40:2} (1980) 688-706, \
Teor.Mat.Fiz.40:194-220,1979 (in Russian)},\ref\algbans{L.D.~Faddeev and L.~Takhtajan, 
Russ. Math. Survey {\it 34} (1979) 11},\ref\algbas{L.D.~Faddeev and L.~Takhtajan, 
J.~Sov.~Math {\it 19} (1982) 1596},\ref\fadalg{L.D.~Faddeev, ``Algebraic aspects of Bethe Ansatz'', Int.~J.~Mod.~Phys. {\bf A}10 (1995)
1845-1878, hep-th/9404013},\ref\fadba{L.D.~Faddeev, ``How algebraic Bethe ansatz works for
integrable
model'',  hep-th/9605187}\ more suggestive. In particular, the raising and the lowering generators of the Yangian algebra are identified with the brane creation and annihilation operators. 
In \bass\ we develop the gauge theory/quantum integrable system correspondence further, by looking at the more exotic gauge theories, coming from higher dimensions via a compactification on a sphere, with a partial twist, or via a localization on a fixed locus of some rotational symmetry. We also discuss the relation of our duality to the familiar story of the classical integrability describing the geometry of the space of vacua of the four dimensional ${\CN}=2$ theories \WitDonagi. 

In \bass\ the Hamiltonians of the quantum integrable system are identified with the operators
of quantum multiplication in the equivariant cohomology of the
hyperk\"ahler quotients, corresponding to the Higgs branches of
our gauge theories. In particular, the length $L$ inhomogeneous $XXX_{1\over 2}$ chain (with  all local spins equal to $\half$) corresponds to the equivariant quantum cohomology of the cotangent bundle $T^{*} Gr(N, L)$ to the Grassmanian $Gr(N, L)$.
This result complements nicely the construction 
of H.~Nakajima and others of the action of the Yangians \gyangian,\mvyangian\ and quantum affine algebras on the classical cohomology and K-theory respectively of certain quiver varieties. Next, \bass\ applies these results to the two dimensional topological field theories. We discuss various twists of our supersymmetric gauge theories. The correlation functions of the
chiral ring operators map to the equivariant intersection indices
on the moduli spaces of solutions to various versions of the
two dimensional vortex equations, with what is mathematically called the Higgs fields taking values in various line bundles (in the case of Hitchin equations the Higgs field is valued in the canonical line bundle).
The main body of \bass\ has essentially shown that all known Bethe ansatz-soluble integrable systems are covered by our correspondence. However, there
are more supersymmetric gauge theories which lead
to the equations \main\  which can be viewed as the deformations of Bethe equations. For example, a four dimensional ${\CN}=2^{*}$ theory compactified on ${\bS}^{2}$
with a partial twist leads to a  deformation
of the non-linear Schr\"odinger system with interesting modular
properties. Another interesting model is the quantum cohomology of the instanton moduli spaces and the Hilbert scheme
of points. 
The long paper \bass\ is reviewed in \ref\basshortcargese{N.~Nekrasov, S.~Shatashvili, 
``Supersymmetric vacua and Bethe ansatz,'' arXiv:hep-th/0901.4744} with the focus on the general nature of
the gauge theory/quantum integrable system
correspondence.

\bigskip

\noindent {\bf Acknowledgments.} We thank V.~Bazhanov, G.~Dvali, L.~Faddeev, S.~Frolov, A.~Gorsky,  K.~Hori, A.~N.~Kirillov,
V.~Korepin, B.~McCoy, M.~Nazarov, A.~Niemi,  A.~Okounkov, E.~Rabinovici, N.~Reshetikhin, S.~J.~Rey,  L.~Takhtajan and A.~Vainshtein and especially A.~Gerasimov and F.~Smirnov, for the discussions. 

The results of this note, as well as those in \bass,  were presented at various conferences and workshops\foot{The IHES seminars and the theoretical physics conference dedicated to the 50th anniversary of IHES  (Bures-sur-Yvette, June 2007, April 2008, June 2008);
 the IAS Workshop on ``Gauge Theory and Representation Theory'' and the IAS seminar (Princeton, November 2007, 2008); the
YITP/RIMS conference ``30 Years of Mathematical Methods
in High Energy Physics
'' in honour of 60th anniversary of Prof. T.~Eguchi (Kyoto, March 2008); the London Mathematical Society lectures at Imperial College  (London, April 2008);  L.~Landau's 100th anniversary
theoretical physics conference (Chernogolovka, June 2008); Cargese Summer Institute (Cargese, June 2008); the Sixth Simons Workshop ``Strings, Geometry and the LHC'' (Stony Brook, July 2008);
the ENS summer institute (Paris, August 2008); the French-Japanese Scientific Forum ''Perspectives in mathematical sciences'', (Tokyo,
October 2008)} and we thank the organizers for the opportunity to present our results. We thank various agencies and institutions\foot{The RTN contract 005104 "ForcesUniverse" (NN and SS),  the ANR grants
ANR-06-BLAN-3$\_$137168 and ANR-05-BLAN-0029-01 (NN), the RFBR grants
RFFI 06-02-17382 and NSh-8065.2006.2 (NN),
 the NSF grant No. PHY05-51164 (NN),  the SFI grants 05/RFP/MAT0036, 08/RFP/MTH1546 (SS) and the Hamilton Mathematics Institute TCD (SS).
Part of research was done while NN visited NHETC at Rutgers University in 2006, Physics and Mathematics Departments of Princeton University in 2007, Simons Center at the Stony Brook University in 2008, KITP at the UC Santa Barbara in 2009, while SSh visited IAS in Princeton in 2007, CERN in 2007 and 2008, Ludwig-Maximilians University in Munich in 2007 and IAS in Jerusalem in 2008.} for supporting this research.

\newsec{Grassmanian and the $XXX$ spin chain.}

\subsec{The gauge theory}

Consider the ${\CN} = (2,2)$ supersymmetric
two dimensional theory with the gauge group $U(N)$, which is a compactification of the $N_{f} = L$, $N_{c} = N$,
four dimensional ${\CN}=2$ theory
on a two-torus.  In four dimensions this theory has
an $SU(N_{f})$ global symmetry group, in addition to the $SU(2)$
non-anomalous and $U(1)$ anomalous (for $N_{f} \neq 2N_{c}$) $R$-symmetry groups. We turn on a Wilson loop
for these symmetry groups (ignoring the anomaly issue for a moment). The condition of unbroken supersymmetry requires these Wilson loops be flat. In the limit of the vanishing two-torus, if these Wilson loops are scaled appropriately, the resulting two dimensional theory has the so-called
twisted mass couplings. The flatness
condition means that the twisted masses belong to the complexification of the Lie algebra of the maximal torus of the global symmetry group. 
Note that there are certain mass couplings, the so-called complex mass terms, which one can turn on already in four dimensions. We first discuss the theory with both complex and twisted masses vanishing.

\subsec{The $T^{*}$Grassmanian sigma model}

Let us analyze the low energy field configurations of the theory whose matter content we just presented. 
The four dimensional theory and its two dimensional reduction have a superpotential 
\eqn\fdsuperpot{
W = {\tr}_{{\bC}^{L}}\, {\tilde Q} {\Phi} Q}
where $Q$ is in the representation $( {\bN}, \bar\bL )$ of $U(N)\times SU(L)$, $\tilde Q$ is in $ ( \bar\bN, \bL )$ of $U(N) \times SU(L)$, and $\Phi$ is in the adjoint of $U(N)$. Recall that $(Q, {\tilde Q})$ form a hypermultiplet, while ${\Phi}$ is a scalar in the vector multiplet of ${\CN}=2$ supersymmetry in four dimensions. 
In two dimensions one gets another complex scalar in the vector multiplet, which we denote by ${\s}$. 

The low energy configurations have vanishing $F$- and $D$-terms. The vanishing of the $F$-terms means:
\eqn\ftvn{Q {
\tilde Q} = 0
\, , \, {\Phi}Q = 0
\, , \, {\tilde Q}{\Phi} = 0}
while the vanishing of the $D$-terms means, in the presence
of the Fayet-Illiopoulos term $r$:
\eqn\dtvn{QQ^{\dagger} - {\tilde Q}^{\dagger}{\tilde Q} + [ {\Phi}, {\Phi}^{\dagger} ] - r \cdot {\bf 1}_{N} = 0}
We recall that the potential of the supersymmetric theory is basically the sum of the absolute squares of the left hand sides of \dtvn\ and \ftvn. 
 
Finally we identify the solutions to
\ftvn\ and \dtvn\ which differ by the $U(N)$ gauge transformations. 
The low energy limit of the gauge theory is a sigma model on the cotangent bundle to the Grassmanian\foot{Let us assume $r > 0$ (the case $r< 0$ is similar, the case $r=0$ is complicated and will not be discussed). By taking the absolute
squares of the norms of \ftvn\ and \dtvn\ we can deduce that $\Phi = 0$, while $Q$ has a maximal rank.  
Then, $(Q, {\tilde Q})$ obeying
\dtvn\ with $\Phi = 0$, define a point in the Grassmanian ${\rm Gr}(N, L)$, as follows:
define a positive definite Hermitian matrix $H = H^{\dagger}$ as the unique square root $H = \left( r{\bf 1}_{N} + {\tilde Q}^{\dagger}{\tilde Q} \right)^{1/2}$. Define:
\eqn\ehq{E = H^{-1}Q\, , \, E^{\dagger} = Q^{\dagger} H^{-1}}
Then $E$ defines an orthonormal
set of $N$ vectors in ${\bC}^{L}$:
\eqn\eed{EE^{\dagger} = {\bf 1}_{N}}
This is our point in the Grassmanian. Now, given $E$, the rest of our data is $F = {\tilde Q} H^{-1}$ obeying $E F = 0$. Indeed,
given $F$, such that $\Vert F \Vert < 1$, we can reconstruct $H$:
\eqn\ffrh{F^{\dagger}F = {\bf 1}_{N} - r H^{-2}}
The matrix $\tilde Q$ defines a point in the cotangent space to the Grassmanian at the point $E$.}. This is a 
hyperk\"ahler manifold, as  required by the ${\CN}=4$ supersymmetry in two dimensions.

The superpotential \fdsuperpot\ is $SU(L)$ invariant (it is actually $U(L)$ invariant, but $U(1)$ is a part of the gauge group). The maximal torus of $SU(L)$ acts as follows:
\eqn\sulqq{\left( {\tilde Q} , Q \right) \mapsto \left( e^{- i m} \,{\tilde Q}\, , \, Q\, e^{i m} \right)}
where
\eqn\mdg{m = {\rm diag} \left( m_{1}, \ldots , m_{L} \right)\ , }
with
$$
\sum_{a=1}^{L} m_{a} = 0
$$
In addition it is invariant under the $U(1)$ symmetry acting as:
\eqn\uoneqfq{e^{+i \upsilon}: \, \left( {\tilde Q}, {\Phi} , Q \right)   \mapsto
\left( e^{-i \upsilon}{\tilde Q}, e^{+2i \upsilon} {\Phi} , e^{-i \upsilon} Q \right)}
These symmetries allow us to turn on the twisted masses. 
We have $L-1$ twisted masses corresponding to the $SU(L)$ symmetry, which we shall parametrize as the mass $u$ which corresponds to the $U(1)$ symmetry \uoneqfq, and $L$ masses $m_{a} $, which are defined up to a common shift, $m_{a} \to m_{a} + {\d}$. Sometimes it is convenient to fixe the gauge by requiring they sum up to zero:
\eqn\mms{\sum_{a=1}^{L} m_{a} = 0\ ,}
but we also shall need other choices.
 
In the sigma model description these symmetries correspond to the isometries of $T^{*}{\rm Gr}(N,L)$. 

\subsec{Supersymmetric ground states}

The main subject of our storty is the space of supersymmetric ground states of the gauge theory. 
In the supersymmetric sigma model description, which is a kind of a Born-Oppenheimer approximation, the ground states correspond to the cohomology of the target space\foot{For the non-compact target spaces one should use some kind of $L^{2}$-cohomology theory. Most of our discussion will be about the theory with twisted masses, where there are no flat directions.}. 

For the cotangent bundle of the Grassmanian the cohomology space is isomorphic to the $N$-th exterior power of the $L$-dimensional vector space:
\eqn\cohgr{H^{*}\left( T^{*}{\rm Gr}(N,L), {\bC} \right) = {\wedge}^{N} {\bC}^{L}}
The isomorphism \cohgr\ is a little bit mysterious since the grading in the cohomology group is not
obvious on the right hand side of \cohgr. The space ${\bC}^{L}$ in the right hand side of \cohgr\ has nothing to do with the space ${\bC}^{L}$ whose $N$-dimensional subspaces are parametrized by the Grassmanian Gr$(N,L)$. Perhaps a bit more geometric description of the cohomology of $T^{*}{\rm Gr}(N,L)$ is via the cohomology of the Grassmanian Gr$(N,L)$ itself. The latter is generated by the Chern classes of the rank $N$ tautological vector bundle $E$. Let
\eqn\bax{{\bQ}(x) = x^{N} - c_{1}(E) x^{N-1} + c_{2}(E) x^{N-2} - \ldots + (-1)^{N} c_{N}(E)}
be the Chern polynomial of $E$. 
The cohomology ring of the Grassmanian ${\rm Gr}(N,L)$ is generated by $c_{1}(E), c_{2}(E), \ldots , c_{N}(E)$. Let us now describe the relations. 
Let $W \approx {\bC}^{L}$ be the topologically trivial vector bundle over Gr$(N,L)$. Let $E^{\perp} = W/E$ be the dual tautological bundle. Let ${\bQ}^{\perp}(x) = x^{L-N} - c_{1}(E^{\perp}) x^{L-N-1} + c_{2}(E^{\perp}) x^{L-N-2} + \ldots + (-1)^{L-N}c_{L-N}(E^{\perp})$
be the Chern polynomial of $E^{\perp}$. Then 
\eqn\baxclass{{\bQ}(x) {\bQ}^{\perp}(x) = x^{L}}
defines the relations among the generators of the cohomology ring\foot{ 
For example, if $N=1$, then the generator is ${\s} = c_{1}(E)$, and the relation \baxclass\ reads:
$(x - {\s} ) {\bQ}^{\perp}(x) = x^{L}$, 
which implies ${\s}^{L} = 0$,
and $c_{k}(E^{\perp}) = (-1)^{k} {\s}^{k}$. The cohomology ring is, in this case, the $L$-dimensional vector space ${\bC}[{\s}]/{\s}^{L}$. 
In general  
the cohomology ring is the space of symmetric polynomials in $N$ variables ${\s}_{1}, \ldots , {\s}_{N}$, subject to the relations 
${\s}_{i}^{L}=0$, $i=1, \ldots , N$. 
A basis in this quotient can be chosen to be the monomial sums:
${\bm}_{\psi} = \sum_{{\pi}\in {\CS}_{N}} \prod_{i=1}^{N} {\s}_{{\pi}(i)}^{{\psi}_{i}+i-N-1}$, 
where $L \geq {\psi}_{1} > {\psi}_{2} > \ldots > {\psi}_{N} \geq 1$. In the quantum cohomology of the Grassmanian itself these relations are modified to ${\s}_{i}^{L} = Q$, see \wittgr.}. With the twisted masses turned on the cohomology $H^{*}(X)$ of the target space is replaced by the equivariant cohomology $H^{*}_{U} (X)$, where $U$ is the global symmetry group.
For $X = {\rm Gr}(N,L)$, $U = SU(L)$ and \baxclass\ is replaced by:
\eqn\baxgrc{{\bQ}(x) {\bQ}^{\perp}(x) = P(x)\ ,}
where $P(x) = \prod_{a=1}^{L} ( x- m_{a})$ is the equivariant Chern polynomial of $W$. For $X = T^{*}$Gr$(N,L)$, $U = SU(L) \times U(1)$ the equivariant cohomology ring is again generated by $c_{k}(E)$, \baxgrc\ with the relations \baxclass\ replaced by:
\eqn\baxclt{P (x-u/2) {\bQ}(x+u) = {\bt}(x) {\bQ}(x)}
where ${\bt}(x)$ is some polynomial of degree $L$.  The spaces of solutions to \baxclt\ and \baxgrc\ are isomorphic, via
the identification:
\eqn\tbx{{\bt}(x) = {\bQ}^{\perp}(x){\bQ}(x+u)}
accompanied by the shift $m_{a} \mapsto m_{a} + u/2$ (note this means the change of gauge \mms\ for the equivariant parameters).

\subsec{Chiral ring}
{}Every two dimensional ${\CN}=(2,2)$ supersymmetric field theory comes with an interesting
algebra: the so-called chiral ring. 
It is the cohomology of one of the nilpotent supercharges of the theory in the space of operators. 

This algebra, thanks to the state-operator correspondence and the standard supersymmetry arguments, is isomorphic, as a vector space, to the space of the supersymmetric ground states, i.e. to the cohomology of the Grassmanian, in our example. 
The ring structure, however, need not be isomorphic to the ring structure of the classical cohomology. It is parametrized, in fact, by the K\"ahler moduli of the target space. The deformed cohomology ring is called the {\it quantum cohomology}, \vafacr,\AandB, and is by now well-studied mathematically \MaxManin. In our case this is a one-parametric deformation of the classical cohomology ring of 
$T^{*}$Gr. 

In addition, if the target space has
isometries, as is the case of the Grassmanian or its cotangent bundle, the theory, and the corresponding chiral algebra can be deformed by the twisted masses. The mathematical counterpart of this theory is called the {\it equivariant Gromov-Witten theory}, it was introduced in \givental. 

The chiral algebra of our theory is, therefore, the $U(1) \times SU(L)$ equivariant quantum cohomology of the cotangent bundle to the Grassmanian $T^{*}{\rm Gr}(N,L)$. 
It can be described using the twisted effective superpotential \bass\ on the ``Coulomb branch'' of the theory. It is a function of the scalars ${\s}$ in the vector multiplet ${\bf\Sigma}$ of the gauge group $U(N)$, obtained by integrating out the massive matter fields:
\eqn\twsppt{\eqalign{& {\tilde W}({\s}) = \sum_{a=1}^{L} \sum_{i=1}^{N}
\left[ \left( {\s}_{i} - m_{a} + u/2 \right) \left( \, {\rm log} \, \left( {\s}_{i} - m_{a} + u/2 \right) - 1\right) + \right. \cr
& \qquad\qquad\qquad\qquad \left. \left( -{\s}_{i} + m_{a} + u/2 \right) \left( {\rm log} \, \left( -{\s}_{i} + m_{a} + u/2 \right) - 1\right) \right] + \cr
& \qquad\qquad\qquad\qquad\qquad\quad + \sum_{i,j=1}^{N} \left( {\s}_{i} - {\s}_{j}  - u \right) \left( \, {\rm log} \left( {\s}_{i} - {\s}_{j}  - u \right) - 1 \right) \cr
& \qquad\qquad\qquad\qquad\qquad\qquad\qquad + 2\pi it \sum_{i=1}^{N} {\s}_{i} \cr}}
where $t =  {\vt \over 2\pi} + i r$ is the linear combination of the Fayet-Illiopoulos term and the theta angle. The quantum cohomology algebra is generated by the symmetric polynomials $c_{k}$, s.t.
\eqn\chrncl{{\bQ}(x) = \prod_{i=1}^{N} ( x - {\s}_{i} ) = x^{N}  - c_{1} x^{N-1} + c_{2} x^{N-2} - \ldots + (-1)^{N} c_{N}}
subject to the relations \main, which become\foot{In the limit $u \to \infty$ with $m_{a}$ finite, upon the redefinition ${\s}_{i} \to {\s}_{i} + u/2$, and the one-loop renormalization 
$t \to t + {L\over 2\pi i} {\rm log} (- u)$ these equations go over to the equivariant quantum cohomology of Gr$(N,L)$}:
\eqn\grbae{\prod_{a=1}^{L} {{\s}_{i}
 - m_{a}  +  u/2 \over {\s}_{i}
 - m_{a}  -  u/2} = e^{2\pi i t}
 \prod_{j\neq i} {{\s}_{i} - {\s}_{j} + u 
 \over  {\s}_{i} - {\s}_{j} - u}}
 The solutions to \grbae\ are the supersymmetric vacua of the gauge theory. 
 
\subsec{The $XXX$ spin chain}

The spin chain is a quantum integrable system, whose commuting Hamiltonians are the generators of an abelian subalgebra of a larger, noncommutative associative algebra, called the Yangian \ref\drinfeldb{V.~Drinfeld, {\it Quantum groups}, Proc. of ICM Berkeley 1986, Academic Press (1986), 798-820}, \ref\sklyaninb{E.~Sklyanin,  {\it Some algebraic structures connected with the Yang-Baxter equation}, Funct. Anal. Appl. {\bf 16} (1982), 27-34}. In our story we shall be dealing with the Yangian $Y({\bf gl}_{2})$ of the ${\bf gl}_{2}$ Lie algebra. The generators of $Y({\bf gl}_{n})$ look like the (negative) loops in ${\bf gl}_{n}$:
\eqn\yna{T_{ij}(x) = {\d}_{ij} + \sum_{p=1}^{\infty} t_{ij}^{(p-1)}x^{-p}}
where $i,j = 1,\ldots , n$. The defining relations of the Yangian are:
\eqn\rtt{[t_{ij}^{(p+1)}, t_{kl}^{(q)}]-
[t_{ij}^{(p)}, t_{kl}^{(q+1)}] = - \left( 
t_{kj}^{(p)}t_{il}^{(q)} - t_{kj}^{(q)}t_{il}^{(p)} \right) }
or, in terms of $T(x)$:
\eqn\rttx{[T_{ij}(x), T_{kl}(y)] = -{ T_{kj}(x)T_{il}(y)-T_{kj}(y) T_{il}(x) \over x - y}}
It follows from \rttx\ that for any diagonal matrix $q = {\rm diag} \left( q_{1}, \ldots , q_{n} \right)$
the following operators
\eqn\trsnf{{\tau}(x, q) = 
\sum_{i=1}^{n} q_{i} T_{ii}(x)}
commute:
\eqn\trsnftr{[ {\tau}(x, q) , {\tau} (y,q ) ] = 0}
Given a representation $\CH$ of the Yangian $Y({\bf gl}_{n})$ one can define a set of commuting quantum Hamiltonians acting in 
$\CH$ as the coefficients of expansion of ${\tau}(x,q)$ at $x = \infty$:
\eqn\tauham{{\tau}(x,q) = H_{0} +
H_{1} x^{-1} + H_{2} x^{-2} + \ldots}
where $H_{0} = \sum_{i=1}^{N} q_{i}$, and 
\eqn\comham{[ H_{l}, H_{m} ] = 0\, , \ l,m= 1, \ldots }
Whether the resulting quantum integrable system is interesting or not, depends on the representation $\CH$. 

The spin chains correspond to $H$ which is obtained by the tensor
product of the so-called evaluation representations. Let $n=2$, and 
let $V_{a} \approx {\bC}^{2}$, $a = 1, \ldots, L$ be $L$ copies of the two dimensional representation of the ${\bf gl}_{2}$ Lie algebra. 
Let
\eqn\spinsp{{\CH} = \bigotimes_{a=1}^{L} V_{a}}
Let $e_{ij}$ be the standard matrix with zeroes everywhere except for the $i$-th column and the $j$-th row, where it has $1$.  Each $V_{a}$ can be viewed as a representation of $Y({\bf gl}_{2})$, via:
\eqn\evalsl{T_{ij}(x)[a] = {\d}_{ij} + {e_{ij} \over x - {\n}_{a}}} 
where ${\n}_{a} \in {\bC}$ is an arbitrary complex parameter. The
representations of the Yangian have these extra parameters, since the commutation relations 
\rttx\ are translation invariant. 
Consider an operator $L_{a}(x)$
which acts in the tensor product
of $V_{a}$ and the auxiliary ${\bC}^{2}$:
\eqn\laxop{{\bL}_{a}(x) = \pmatrix{T_{11}(x)[a] & T_{12}(x)[a] \cr
T_{21}(x)[a] & T_{22} (x)[a] \cr} = {\bf 1} + {1\over x - {\n}_{a}}
\pmatrix{ e_{11}[a] & e_{12}[a] \cr
e_{21}[a] & e_{22}[a] \cr} \in \, {\rm End} \left( V_{a} \otimes {\bC}^{2} \right) }
(the index $[a]$ should not be confused with the superscript $(p)$
in the definition of the expansion modes of the Yangian generators). 
Note that:
\eqn\eper{\pmatrix{ e_{11} & e_{12} \cr
e_{21} & e_{22} \cr} = \pmatrix{1 & 0 & 0 & 0 \cr
0 & 0 & 1 & 0 \cr
0 & 1 & 0 & 0 \cr
0 & 0 & 0 & 1 \cr} = P_{12}}
is the permutation matrix acting
in ${\bC}^{2} \otimes {\bC}^{2}$. 
Now let us define
the operators $T_{ij}(x) \in {\rm End} ({\CH})$ via:
\eqn\trsnfr{\pmatrix{T_{11}(x) & T_{12} (x) \cr
T_{21}(x) & T_{22}(x) \cr} = {\bL}_{1}(x) {\bL}_{2}(x) \ldots {\bL}_{L}(x)} 
where the product is the matrix product in the auxiliary space ${\bC}^{2}$. More precisely, we should view the operators $e_{ij}$ in \laxop\ as acting in $\CH$, as:
\eqn\eija{\eqalign{
& \qquad\qquad\qquad\ {\rm a'th}_{\downarrow}\, {\rm factor} \cr
& e_{ij}[a] = {\bf 1} \otimes \ldots \otimes e_{ij} \otimes \ldots \otimes {\bf 1} \cr}}
The commutation relations \rttx\ can be written in the so-called R-matrix form, which makes it obvious that \trsnfr\ defines a representation of $Y({\bf gl}_{2})$ in $\CH$. Since the scaling of $q$ by an overall factor simply multiplies ${\tau}(x,q)$ by the same factor, one can set $q_{1} = 1$ without much loss of the generality. 

The parameter $Q = q_{2}/q_{1}$
is called the twist parameter. The parameters ${\n}_{a}$ are called the impurities. The translation invariance of \rttx\ again implies that the shift of all ${\n}_{a}$'s by the same amount leads to physically equivalent system. 

\subsec{Special cases and limits}

The Heisenberg spin chain corresponds to ${\n}_{a} = 0$, $Q = 1$.

The higher spins can be obtained by arranging the impurities in a special way and then restricting onto an irreducible submodule. In this way one can get the local spins to be in arbitrary representations of ${\bf sl}_{2}$. 
This is done by choosing, e.g.
\eqn\spimp{\{ {\n}_{a} \}  = \{ {\bar\n}_{A} + k_{A}\}\, , \  k_{A} = -{\hat s}_{A}, -{\hat s}_{A}+1, \ldots , {\hat s}_{A}-1, {\hat s}_{A}\,
 , }where ${\hat s}_{A} = s_{A} - {\half}$, $2s_{A} \in {\bZ}_{+}$, $A = 1 , \ldots , {\ell}$. Thus, one can view the spin chain of length ${\ell}$ with the local spins $s_{1}, \ldots , s_{\ell}$
as a subsystem in the spin $\half$ chain of the length $L = 
\sum_{A} 2s_{A}$.    

Let us now assume that all local spins are taken to infinity, and at the same time the number of spin sites $L$ is also sent to infinity, so that
\eqn\sac{\sum_{a=1}^{L} {1\over s_{a}} = {1\over c} \ {\rm finite}}
In this limit the spin chain goes over to the non-linear Schr\"odinger system, NLS. 
 
Now keep $L$ and $s_{a}$ finite, but take ${\n}_{a} \to \infty$, instead. In this case the spin chain goes over to the so-called Gaudin system. 

\subsec{Bethe ansatz} 

Solution of the quantum integrable system consists of finding the common eigenvectors of the commuting Hamiltonians. 
There is the following ansatz for the  spin chains eigenfunctions:
\eqn\baet{{\bf\Psi}_{\l} = T_{12}({\l}_{1})     T_{12}({\l}_{2})   \ldots  
 T_{12}({\l}_{N}) \, {\Omega}}
 where 
 $\Omega \in \CH$ is the highest weight vector:
 \eqn\baehw{\eqalign{& T_{21} (x)  \, {\Omega} = 0\cr
 & T_{11}(x) \, {\Omega} = r_{+}(x) \, {\Omega}\cr
 &  T_{22}(x) \, {\Omega} = r_{-}(x) \, {\Omega}\cr}}
     where $r_{\pm}(x)$ are the easily computed rational functions:
     \eqn\rpm{r_{\pm}(x) = {P(x \pm \half)  \over P(x)} \, , \ P(x) = \prod_{a=1}^{L} \, \left( x - {\n}_{a}  \right)}
     In fact, the vector $\Omega$ is just the tensor product of the spin down states over all $L$ factors $V_{a}$.  The vector \baet\ is the eigenvector of ${\tau}(x,q)$ for all $x$ iff $\l$'s solve the following system of equations, the {\it Bethe 
equations}:
\eqn\bae{\prod_{a=1}^{L} {{\l}_{i} - {\n}_{a} + {\half} \over {\l}_{i} - {\n}_{a} - {\half}} = Q \prod_{j \neq i} {{\l}_{i} - {\l}_{j} + 1 \over {\l}_{i} - {\l}_{j} - 1}} 
Equivalently, the polynomial, the so-called Baxter operator, \baxter:
\eqn\baxop{{\bQ}(x) = \prod_{i=1}^{N} ( x - {\l}_{i} )}
has to obey the following difference equation:
\eqn\baxeq{r_{+}(x) {\bQ}(x - 1) + Q\, r_{-}(x) {\bQ}(x+1) = {\e} (x) {\bQ}(x)}
where ${\e}(x)P(x)$ is a degree $L$ polynomial in $x$, to be determined from \baxeq\ at the same time as ${\bQ}(x)$. The equations \bae\  imply that the left hand side of \baxeq\ vanishes at $x = {\l}_{i}$ and therefore is divisible by ${\bQ}(x)$. The rational function ${\e}(x)$ is the eigenvalue of the transfer matrix ${\tau}(x,q)$. 

Finally, the equations \bae\ can be interpreted as the critical point equations for the YY function:
\eqn\yngfnc{\eqalign{& Y({\l}) = \sum_{a=1}^{L} \sum_{i=1}^{N}
\left[ \left( {\l}_{i} - {\n}_{a} + {\half} \right) \left( \, {\rm log} \, \left( {\l}_{i} - {\n}_{a} + {\half} \right) - 1\right) + \right. \cr
& \qquad\qquad\qquad \left. \left( -{\l}_{i} + {\n}_{a} + {\half} \right) \left( {\rm log} \, \left( -{\l}_{i} + {\n}_{a} + {\half} \right) - 1\right) \right] + \cr
& \qquad\qquad\qquad\qquad \sum_{i,j=1}^{N} \left( {\l}_{i} - {\l}_{j}  - 1 \right) \left( \, {\rm log} \left( {\l}_{i} - {\l}_{j}  - 1 \right) - 1 \right) \cr}}

\subsec{Dictionary}

We can now formulate the precise dictionary. The operators \chrncl\ of the gauge theory map to Baxter
operator \baxop, upon the rescaling, ${\s}_{i} = {\l}_{i}u$, and make it the lowest component of the superfield,
\eqn\baxops{{\bQ}(x) = {\rm det} \left( x - {\bf\Sigma} \right)} 
the twisted superpotential is identified with the YY function, 
\eqn\wyy{{\tilde W} ({\s}; m, s) = u\, Y \left( {\s}/u ; {\n}, s \right)\ , }
while the supersymmetric vacua are mapped one-to-one to the Bethe eigenstates, provided these form a complete set. 
The twisted masses corresponding to the $SU(L)$ symmetry are the impurities, $m_{a} = {\n}_{a}u$ while the twisted mass corresponding to the $U(1)$ symmetry \uoneqfq\ sets the scale of the twisted masses. The instanton factor is mapped to the twist parameter:
\eqn\insttw{e^{2\pi i t} = Q}
The roots of Drinfeld polynomial $P(x)$ are the masses of the fundamental hypermultiplets. Baxter's equations \baxeq\ become the Ward-like identities of the chiral ring:
\eqn\baxeqq{ \left\langle \, \left[ \, 
P ( x + u/2 ) {{\bQ} (x - u )\over {\bQ} (x)} + e^{2\pi it}\, 
P ( x - u/2 ) {{\bQ} ( x + u ) \over 
{\bQ} (x) } \, \right]_{-} \, \right\rangle = 0}
where $[ \ldots ]_{-}$ denotes a negative in $x$ part in the expansion near $x = \infty$. We have established \baxeqq\ via the correspondence with the spin chain, yet it would be nice to derive \baxeqq\ by a direct gauge theory argument. 

In the string theory realization $u$ is mapped to the topological string coupling constant. The identification of the K\"ahler parameter $t$ with the element of the Cartan  subalgebra of ${\bf sl}_{2}$ (and the analogous identification in the higher rank case) is a particular case of the famous string realization of McKay  duality between the ADE singularities and the Lie groups.

The $XXZ$ generalization of the spin chain maps to the $2+1$ gauge theory compactified on a circle, the $XYZ$ spin chain and the $8$-vertex models \baxter, \ref\bazhmang{V.~V.~Bazhanov, V.~V.~Mangazeev, 
``Analytic theory of the eight-vertex model,''
 arXiv:hep-th/0609153, \np{775}{2007}{225-282}} are dual to the four dimensional gauge theory compactified on a two-torus.

The lack of irreducibility of the Yangian representation which occurs at the special arrangement \baekr\ of impurities \spimp\ translates to the vanishing of the mass gap\foot{The massive theories lead to the semi-simple Frobenius manifolds and the diagonalizable quantum cohomology ring \dubrovintt} at some of the vacua of the gauge theory \bass. For the special values of the twisted masses there appear extra massless particles in the spectrum, typically mesons. The theory can be further deformed by the superpotential terms (corresponding to the complex masses), and the irreducible representation of the Yangian can be extracted, by lifting the flat branches. Similar phenomena occur in other models, at $Q \to 0$ in the $XXX$ model \nazarov, in the $XXZ$ model with $e^{i u}$ being a root of unity, in the six-vertex model \ref\mccoy{K.~Fabricius, B.~McCoy,``Completing Bethe's equations at roots of unity'', arXiv:cond-mat/0012501 \semi
K.~Fabricius, B.~McCoy, ``Bethe's equation is incomplete for the XXZ model at roots of unity'', arXiv:cond-mat/0009279}, etc.

\subsec{Beyond the known systems}
 
 The correspondence with the supersymmetric gauge theories opens new doors both for the
 quantum integrable systems and for the gauge theories. 
 
For example, in \bass, \basshortcargese\ we study four dimensional ${\CN}=2$ gauge theories compactified on a two-sphere, with a partial twist. The effective twisted superpotential of the resulting two-dimensional theory gives rise to a (modular-covariant) deformation of the Bethe equations of the non-linear Schr\"odiner system. 

Another very exciting direction
of research involves attempting to lift the correspondence between the quantum integrable system and the gauge theory beyond the vacuum sector of the latter. It is conceivable that the
Yangian, quantum affine, or elliptic quantum algebra symmetry of the vacuum sector are the symmetries of the full, interacting quantum field theory, once we combine the theories with different rank gauge groups.
Viewing the rank of a gauge group as a superselection sector parameter is natural in the noncommutative field theory context and even more natural
in the context of string theory.

\bigskip
\centerline{* * *}
\centerline{All these questions are discussed in greater detail in \bass. }

\footatend\vfill\supereject\immediate\closeout\rfile\writestoppt
\baselineskip=14pt\centerline{{\bf References}}\bigskip{\frenchspacing%
\parindent=20pt\escapechar=` \input refs.tmp\vfill\eject}\nonfrenchspacing
\bye